\documentclass{PoS}
\usepackage[pass]{geometry} 
\usepackage[T1]{fontenc}
\usepackage{color, braket, bm, subfig, mathrsfs,mathtools, wasysym, amsmath, amsfonts, amssymb, graphicx, slashed, cite}
\newcommand{\tex}{\textsc{t\kern -.12em\lower.4ex\hbox{e}\kern-.1em x}}
\newcommand{\col}{\textcolor{blue}}

\title{Looking for New Naturally Aligned Higgs Doublets at the LHC}

\ShortTitle{Looking for New Naturally Aligned Higgs Doublets at the LHC}

\author{P. S. Bhupal Dev \\
Consortium for Fundamental Physics,
  School of Physics and Astronomy, University of Manchester, Manchester M13 9PL, United Kingdom.\\
Physik Department T30d,
Technische Universit\"{a}t M\"{u}nchen, James-Franck-Stra\ss e 1, 85748 Garching, Germany. \\
        E-mail: \email{bhupal.dev@tum.de}}

\author{\speaker{Apostolos Pilaftsis} \\ 
       Consortium for Fundamental Physics,
  School of Physics and Astronomy, University of Manchester, Manchester M13 9PL, United Kingdom.\\
CERN, Physics Department, Theory Division, CH-1211 Geneva 23, Switzerland. \\
       E-mail: \email{Apostolos.Pilaftsis@manchester.ac.uk}}

\abstract{Since the current LHC Higgs data suggest the couplings
       of the observed 125 GeV Higgs boson to be close to the Standard
       Model (SM) expectations, any extended Higgs sector must lead to
       the so-called SM {\em alignment limit}, where one of the Higgs
       bosons behaves exactly like that of the SM. In the context of
       the Two Higgs Doublet Model (2HDM), this alignment is often
       associated with either decoupling of the heavy Higgs sector or
       accidental cancellations in the 2HDM potential. We present a
       novel symmetry justification for `natural' alignment without
       necessarily decoupling or fine-tuning. We show that there exist
       only {\em three} different symmetry realizations of the natural
       alignment scenario in 2HDM. We analyze new collider signals for
       the heavy Higgs sector in the natural alignment limit, which
       dominantly lead to third-generation quarks in the final state
       and can serve as a useful observational tool during the Run-II
       phase of the LHC.}

\FullConference{18th International Conference From the Planck Scale to the Electroweak Scale \\
        25-29 May 2015\\
        Ioannina, Greece }

\begin{document}

\section{Introduction}\label{sec:1}
The discovery of a Higgs boson in the Run-I phase of the LHC~\cite{Aad:2012tfa} provides the first experimental evidence for the Higgs mechanism~\cite{Higgs:1964pj} as the standard theory of electroweak symmetry breaking (EWSB). The
measured  couplings of  the  discovered Higgs  boson are  remarkably consistent 
  with  those   predicted   by  the   Standard Model (SM)~\cite{coup}. However,  the
possibility of an extended Higgs sector, as suggested by various  well-motivated  new-physics scenarios, such as supersymmetry, is still allowed by the current  experimental  data. 

Here we consider the simplest extension of the standard Higgs mechanism, namely the Two Higgs Doublet Model~(2HDM)~\cite{review}, where the SM Higgs doublet is supplemented by another isodoublet with hypercharge $Y=1$. 
In the doublet field space $\Phi_{1,2}$, where $\Phi_i = (\phi_i^+, \phi_i^0)^{\sf T}$, the  general 2HDM  potential reads 
\begin{align}
  \label{pot}
V \ = \ & -\mu_1^2(\Phi_1^\dag \Phi_1) -\mu_2^2 (\Phi_2^\dag \Phi_2) 
-\left[m_{12}^2 (\Phi_1^\dag \Phi_2)+{\rm H.c.}\right]  \nonumber \\ &
+\lambda_1(\Phi_1^\dag \Phi_1)^2+\lambda_2(\Phi_2^\dag \Phi_2)^2 
 +\lambda_3(\Phi_1^\dag \Phi_1)(\Phi_2^\dag \Phi_2) 
 +\lambda_4(\Phi_1^\dag \Phi_2)(\Phi_2^\dag \Phi_1) 
\nonumber \\
& 
+\left[\frac{\lambda_5}{2}(\Phi_1^\dag \Phi_2)^2
+\lambda_6(\Phi_1^\dag \Phi_1)(\Phi_1^\dag \Phi_2) 
+\lambda_7(\Phi_1^\dag \Phi_2)(\Phi_2^\dag \Phi_2)+{\rm H.c.}\right],
\end{align}
which contains  {\em four} real mass  parameters $\mu_{1,2}^2$, Re$(m^2_{12})$,
Im$(m^2_{12})$,  and {\em ten}  real  quartic couplings  $\lambda_{1,2,3,4}$,
Re($\lambda_{5,6,7})$,  and Im($\lambda_{5,6,7}$). Thus,
the   vacuum   structure   of   the   general  2HDM   can   be   quite
rich~\cite{Battye:2011jj}, as compared to the SM. 

The quark-sector Yukawa Lagrangian in the general 2HDM is given by 
\begin{align}
-{\cal L}^q_Y \  = \  \bar{Q}_L(h_1^d \Phi_1+ h_2^d\Phi_2)d_R \: +
\: \bar{Q}_L(h_1^u \widetilde{\Phi}_1+ h_2^u \widetilde{\Phi}_2)u_R \; ,
\label{yuk}
\end{align}
where $\widetilde{\Phi}_i={\rm i}\sigma^2\Phi_i^*$ ($\sigma^2$ being the second Pauli matrix) are the isospin conjugates of $\Phi_i$, $Q_L=(u_L,d_L)^{\sf T}$ are the $SU(2)_L$ quark doublets and $u_R,d_R$ are 
right-handed quark singlets. To avoid potentially large 
flavor-changing neutral current processes at the tree level induced by the Yukawa interactions in~\eqref{yuk}, one imposes a discrete $Z_2$ symmetry~\cite{review} under which 
\begin{align}
\Phi_1 \to -\Phi_1,\quad \Phi_2\to \Phi_2, \quad u_{Ra}\to u_{Ra},\quad 
d_{Ra}\to d_{Ra}~~{\rm or}~~ d_{Ra}\to -d_{Ra} \; ,
\label{discrete}
\end{align}
($a=1,2,3$ being the quark family index) so that only $\Phi_2$ gives mass to up-type quarks, and 
only $\Phi_1$ or only $\Phi_2$ gives mass to down-type quarks. 
The $Z_2$ symmetry~\eqref{discrete} is satisfied  by four  discrete  
choices of  tree-level Yukawa  couplings between the Higgs doublets and  SM fermions, which are  known as the Type I, II, X (lepton-specific) and Y (flipped) 2HDMs~\cite{review}. Global fits to the LHC Higgs data (see e.g.,~\cite{fit1, Chowdhury:2015yja}) suggest that all four types of 2HDMs with natural flavor conservation are constrained to
lie close to the so-called SM {\it alignment limit}~\cite{Georgi:1978ri, Gunion:2002zf, Carena:2013ooa, Dev:2014yca, Das:2015mwa}, where the mass
eigenbasis of the  CP-even scalar sector aligns with  the SM gauge
eigenbasis.  

The SM  alignment is  often associated with  the decoupling  limit, in
which all the non-standard Higgs bosons are assumed to be much heavier
than the electroweak scale so that the lightest CP-even scalar behaves
just  like the  SM  Higgs  boson.  The  alignment  limit  can also  be
achieved,  without decoupling~\cite{Chankowski:2000an,  Gunion:2002zf,
  Carena:2013ooa}, but  for small $\tan\beta$ values,  this is usually
attributed  to accidental  cancellations  in the  2HDM potential.   We
present a novel  symmetry argument to naturally  justify the alignment
limit~\cite{Dev:2014yca}, independently of the kinematic parameters of
the theory, such as the heavy Higgs mass and~$\tan\beta$. In Section~\ref{sec:3}, we show 
that there exist  only {\em  three} possible  symmetry realizations  of the
scalar  potential  which  predict natural  alignment, as defined in Section~\ref{sec:2}.  
We  explicitly
analyze  the  simplest  case,  namely, the  Maximally  Symmetric  2HDM
(MS-2HDM), which realizes a SO(5) symmetry in the bilinear field basis
to be discussed  in Section~\ref{sec:3}. We show that  the renormalization group (RG)  effects  due   to  the  hypercharge  gauge   coupling  $g'$  and
third-generation  Yukawa  couplings,  as well  as  soft-breaking  mass
parameter $m^2_{12}$, induce relevant deviations from the SO(5) limit,
which  lead to  distinct predictions  for  the Higgs  spectrum of  the
MS-2HDM (see Section~\ref{sec:4}).  In particular, a striking feature of the SO(5) limit is that
the  heavy Higgs  sector is  predicted to  be {\em  quasi-degenerate},
apart from being {\em gaugephobic}, which  is a generic feature in the
alignment limit.  Moreover, the current experimental constraints force
the heavy  Higgs sector to  lie above  the top-quark threshold  in the
MS-2HDM. Thus, the  dominant collider signal for  this sector involves
final  states with  third-generation quarks.   We make  a parton-level
study of  some of  these signals (see Section~\ref{sec:5}),  which can be  useful for  the heavy
Higgs searches in the ongoing Run-II phase of the LHC.


\section{Natural Alignment Condition}\label{sec:2}
For simplicity, we consider the CP-conserving 2HDM, but our results can be easily generalized to the CP-violating case. 
After EWSB by the vacuum expectation values (VEVs) $v_{1,2}$ of the two scalar fields $\Phi_{1,2}$, there  are five physical scalar mass eigenstates: two CP-even ($h,H$), one
CP-odd ($a$)  and two charged  ($h^\pm$) scalars. The corresponding mass     eigenvalues     are     given by~\cite{Haber:1993an}
\begin{subequations}
\begin{align}
M^2_{h^\pm} \ & = \ \frac{m_{12}^2}{s_\beta
  c_\beta}-\frac{v^2}{2}\left( \lambda_4+ \lambda_5\right)
 +\frac{v^2}{2s_\beta c_\beta}\left( \lambda_6
c_\beta^2+ \lambda_7 s_\beta^2\right), \qquad 
M_a^2 \   = \  M^2_{h^\pm}+\frac{v^2}{2}\left( \lambda_4 - 
\lambda_5\right), \label{mass1} \\ 
M_H^2 \ & = \ \frac{1}{2}\left[(A+B)-\sqrt{(A-B)^2+4C^2}\right], \qquad 
M_h^2 \  = \ \frac{1}{2}\left[(A+B)+\sqrt{(A-B)^2+4C^2}\right], \label{mass3}
\end{align} 
\end{subequations}
where we have used the short-hand notations $c_\beta\equiv \cos\beta$ and $s_\beta\equiv \sin\beta$ with $\tan\beta=v_2/v_1$ and    
\begin{subequations}
\begin{align}
A \ & = \ M_a^2s_\beta^2+v^2\left( 2\lambda_1 c_\beta^2+ 
\lambda_5 s^2_\beta+2 \lambda_6 s_\beta
  c_\beta\right), \\ 
B \ & = \ M_a^2c_\beta^2+v^2\left( 2\lambda_2 s_\beta^2+ 
\lambda_5 c^2_\beta + 2\lambda_7 s_\beta c_\beta
  \right) , \\ 
C \ & = \ -M_a^2 s_\beta c_\beta + v^2\left( \lambda_{34}s_\beta c_\beta
  + \lambda_6 c^2_\beta + \lambda_7 s^2_\beta
  \right) . 
\end{align}
\end{subequations}
with $\lambda_{34}=\lambda_3+\lambda_4$. The mixing between the mass eigenstates in the
CP-odd  and charged  sectors is  governed by the angle $\beta$, whereas in the CP-even sector, it is governed by the angle $\alpha=(1/2)\tan^{-1}[2C/(A-B)]$. The SM Higgs field can be identified as the linear combination  
\begin{eqnarray}
H_{\rm SM} \ = \  
H\cos(\beta-\alpha)+h\sin(\beta-\alpha) \; . 
\label{HSM}
\end{eqnarray}
Thus, the  couplings of  $h$ and $H$  to the SM gauge bosons
$V=W^\pm, Z$ with  respect to the SM Higgs  couplings $g_{H_{\rm SM}VV}$ will be respectively $\sin{(\beta-\alpha)}$ and $\cos{(\beta-\alpha)}$.  
The {\em SM alignment limit} is defined as $\alpha\to \beta$ (or $\alpha\to \beta-\pi/2$) when $H$ ($h$) couples to vector bosons exactly as in the SM, whereas $h$ ($H$) becomes {\em gaugephobic}.  For concreteness, we will take the alignment limit as $\alpha\to \beta$.

To derive the alignment condition, we rewrite the CP-even scalar mass matrix as
\begin{align}
M^2_{S} \  & = \ \left(\begin{array}{cc} A& C \\ C & B \end{array}\right) 
\  = \ \left(\begin{array}{cc}
c_\beta & -s_\beta \\ 
s_\beta & c_\beta 
\end{array}\right) \left(\begin{array}{cc}
\widehat{A} & \widehat{C} \\
\widehat{C} & \widehat{B}
\end{array}\right)
\left(\begin{array}{cc}
c_\beta & s_\beta \\ 
-s_\beta & c_\beta 
\end{array}\right)\; , \label{align2} \\
{\rm where} \quad \widehat{A} & \ = \  2v^2 \Big[ c_\beta^4 \lambda_1 
+ s_\beta^2 c_\beta^2 \lambda_{345}  
+ s_\beta^4 \lambda_2\:  +\: 2 s_\beta c_\beta \Big( c^2_\beta \lambda_6 +
s^2_\beta \lambda_7\Big)\Big]\; , \nonumber \\
\widehat{B} & \ = \  M_a^2\: +\: \lambda_5 v^2\: +\: 2 v^2 
\Big[ s^2_\beta c^2_\beta
  \Big(\lambda_1+\lambda_2-\lambda_{345}\Big)\:
-\: s_\beta c_\beta \Big(c^2_\beta - s^2_\beta\Big) \Big(\lambda_6 
- \lambda_7\Big) \Big]\; , \label{bhat} \\
\widehat{C} & \  =   \
v^2 \Big[ s^3_\beta c_\beta \Big( 2\lambda_2-\lambda_{345}\Big) - 
c^3_\beta s_\beta \Big(2\lambda_1- \lambda_{345}\Big) + c^2_\beta
\Big( 1 - 4 s^2_\beta \Big) \lambda_6 
+ s^2_\beta \Big( 4 c^2_\beta - 1\Big) \lambda_7 \Big],  \nonumber 
\end{align}
and  we have used the short-hand notation $\lambda_{345} \equiv  \lambda_3  +
\lambda_4    +    \lambda_5$.      
Evidently, the SM alignment limit $\alpha \to \beta$ is obtained when 
$\widehat{C} =  0$ in~\eqref{align2}~\cite{Gunion:2002zf}. This yields the quartic equation
\begin{eqnarray}
\lambda_7 \tan^4\beta -  (2\lambda_2-\lambda_{345})\tan^3\beta + 3(\lambda_6-\lambda_7)\tan^2\beta + 
(2\lambda_1-\lambda_{345})\tan\beta - \lambda_6 \ = \ 0 \; .
\label{align-gen}
\end{eqnarray}
For {\em natural} alignment, (\ref{align-gen}) should be satisfied for {\it any} value of $\tan\beta$, which requires the coefficients of the polynomial in $\tan\beta$ to vanish identically~\cite{Dev:2014yca}.  This implies 
\begin{eqnarray}
2\lambda_1 \ = \ 2\lambda_2 \ = \ \lambda_{345}\;, \qquad \lambda_6 \ = \ \lambda_7 \ = \ 0\; . 
\label{alcond}
\end{eqnarray}
In particular, for $\lambda_6 = \lambda_7 = 0$ as in the $Z_2$-symmetric 2HDMs, (\ref{align-gen}) has a solution 
\begin{eqnarray}
  \label{tanb}
\tan^2\beta\ =\ \frac{2\lambda_1 - \lambda_{345}}{2\lambda_2 -
\lambda_{345}}\ >\ 0 \; ,
\end{eqnarray}
independent of $M_a$. After some algebra, the simple solution (\ref{tanb}) to our {\em natural alignment condition} (\ref{align-gen}) can be shown to be equivalent to that derived in~\cite{Carena:2013ooa}.

\section{Symmetry Classifications of the 2HDM Potential}\label{sec:3}
 
To identify all  accidental symmetries of the 2HDM potential~\eqref{pot},
it   is   convenient    to  work in the {\em bilinear scalar field formalism}~\cite{Maniatis:2006fs} by introducing  an  8-dimensional  complex multiplet ${\bm \Phi} \equiv (\Phi_1, \Phi_2, \widetilde{\Phi}_1, \widetilde{\Phi}_2)^{\sf T}$~\cite{Battye:2011jj,Nishi:2011gc,Pilaftsis:2011ed}. 
In terms of the ${\bm \Phi}$-multiplet, we define  a {\em   null}   6-dimensional   Lorentz   vector  $R^A\ \equiv\ {\bm \Phi}^\dag \Sigma^A {\bm \Phi}$, 
where  $A=0,1,...,5$  and  the  six $8\times  8$-dimensional  matrices
$\Sigma^A$ may be expressed in terms of the three Pauli matrices $\sigma^{1,2,3}$ and the identity matrix $\bm{1}_{2\times 2}\equiv \sigma^0$, as follows:
\begin{eqnarray}
&& \Sigma^{0,1,3} \ = \ \frac{1}{2}\sigma^0 \otimes \sigma^{0,1,3} \otimes
  \sigma^0, \qquad   
\Sigma^2 \ = \ \frac{1}{2}\sigma^3 \otimes \sigma^2 \otimes \sigma^0, \nonumber\\
&& \Sigma^4 \ = \ -\frac{1}{2}\sigma^2 \otimes \sigma^2 \otimes \sigma^0, \qquad
\Sigma^5 \ = \ -\frac{1}{2}\sigma^1 \otimes \sigma^2 \otimes \sigma^0. 
\end{eqnarray}
Note that the bilinear  field space spanned by the  6-vector $R^A$ realizes
an {\em orthochronous} ${\rm SO}(1,5)$ symmetry group~\cite{Battye:2011jj, Pilaftsis:2011ed}.

\begin{table}[t!]
\begin{center}
\begin{tabular}{c|ccccccccc}\hline
symmetry & $\mu_1^2$ & $\mu_2^2$ & $m^2_{12}$ & $\lambda_1$ & $\lambda_2$ & $\lambda_3$ & $\lambda_4$ & ${\rm Re}(\lambda_5)$ & $\lambda_6=\lambda_7$ \\ \hline
$Z_2\times$ O(2) & - & - & Real & -& -& -& -& -& Real\\
$(Z_2)^2\times $SO(2) & -&  - & 0 & - & -&  - & - & - & 0 \\
$(Z_2)^3\times $O(2) & -&  $\mu_1^2$ & 0 & - &  $\lambda_1$ &  - & - & - & 0 \\
O(2) $\times $O(2) & -&  - & 0 & - & -&  - & - & 0 & 0 \\
\col{$Z_2\times $ [O(2)]$^2$} & -& $\mu_1^2$ & 0 & - & $\lambda_1$ &  - & - & $2\lambda_1-\lambda_{34}$ & 0 \\
\col{O(3)$\times $O(2)} & -&  $\mu_1^2$ & 0 & - & $\lambda_1$ &  - & $2\lambda_1-\lambda_3$  & 0 & 0 \\
SO(3) & -&  - & Real & - & - &  - & -  & $\lambda_4$ & Real \\
$Z_2\times $O(3) & - &  $\mu_1^2$ & Real & - & $\lambda_1$ &  - & -  & $\lambda_4$ & Real \\
$(Z_2)^2\times $SO(3) & -&  $\mu_1^2$ & 0 & - & $\lambda_1$ &  - & -  & $\pm \lambda_4$ & 0 \\
O(2)$\times $O(3) & -&  $\mu_1^2$ & 0 & - & $\lambda_1$ &  $2\lambda_1$ & -  & 0 & 0 \\
SO(4) & -&  - & 0 & - & - &  - & 0  & 0 & 0 \\
$Z_2\times $O(4) & -&  $\mu_1^2$ & 0 & - & $\lambda_1$ &  - & 0  & 0 & 0 \\
\col{SO(5)} & -&  $\mu_1^2$ & 0 & - & $\lambda_1$ & $2\lambda_1$ & 0 &
                                                                       0 & 0 \\ \hline 
\end{tabular}
\end{center}
\caption{Relations between the parameters of the $U(1)_Y$-invariant
  2HDM potential~(1.1) for the 13 accidental
  symmetries~\cite{Pilaftsis:2011ed} 
in a diagonally reduced basis,
  where ${\rm Im} (\lambda_5) = 0$ and $\lambda_6=\lambda_7$.}  
\label{tab1}
\end{table}

In terms of  the null-vector $R^A$, the  2HDM potential (\ref{pot}) 
takes on a simple quadratic form:
\begin{equation}
  \label{potR}
V\ =\ -\, \frac{1}{2}\, M_A\,R^A\: + \: \frac{1}{4}\, L_{AB}\, R^A R^B\; ,
\end{equation}
where $M_A$  and $L_{AB}$ are ${\rm SO}(1,5)$  constant `tensors' that
depend  on the  mass parameters  and quartic  couplings given in~\eqref{pot}   
 and    their   explicit    forms    may   be    found
in~\cite{Pilaftsis:2011ed,Maniatis:2007vn}.
Requiring   that    the   SU(2)$_L$   gauge-kinetic    term   of   the 
{\boldmath $\Phi$}-multiplet  remains canonical restricts  the allowed
set of  rotations from SO(1,5)  to SO(5),  where only  the  spatial components
$R^I$ (with $I=1,...,5$) transform and the zeroth component $R^0$
remains invariant. Consequently, in  the absence of  the hypercharge
gauge coupling and fermion Yukawa couplings, the maximal symmetry
group of the 2HDM is $G^R_{\rm  2HDM} = {\rm SO(5)}$. Including all proper, improper 
and semi-simple subgroups of SO(5), the accidental symmetries for the 2HDM 
potential were completely classified in~\cite{Battye:2011jj, Pilaftsis:2011ed}, as shown in Table~\ref{tab1}. Here we have used a diagonally reduced basis~\cite{Gunion:2005ja}, where ${\rm Im}(\lambda_5)=0$ and $\lambda_6=\lambda_7$, thus reducing the number of independent quartic couplings to seven. Each of the symmetries listed in Table~\ref{tab1} leads to certain constraints on the mass and/or coupling parameters.  From Table~\ref{tab1}, we find that 
there are {\it only} three symmetries, namely $Z_2\times [{\rm O(2)}]^2$, O(3)$\times$ O(2), and  SO(5), which satisfy the natural alignment condition given by~\eqref{alcond}.\footnote{In Type-I 2HDM, there exists an additional possibility of realizing an exact Z$_2$ symmetry~\cite{Deshpande:1977rw} which leads to an exact alignment, i.e. in the context of the so-called inert 2HDM~\cite{Barbieri:2006dq}.} 
In what follows, we focus on the simplest realization of the SM alignment, namely, the MS-2HDM based on  the SO(5) group~\cite{Dev:2014yca}. 

\section{Maximally Symmetric 2HDM}\label{sec:4}
From Table~\ref{tab1}, we see that the maximal symmetry  group in the bilinear field space is SO(5), in which case the parameters of the 2HDM potential~\eqref{pot} satisfy the following relations:
\begin{align}
& \mu_1^2 \ = \ \mu_2^2\; , \quad m^2_{12} \ = \ 0\; , \quad \nonumber \\
& \lambda_2 \ = \ \lambda_1\; , \quad 
 \lambda_3  \ = \ 2\lambda_1\; , \quad 
\lambda_4 \ = \  {\rm Re}(\lambda_5) \  = \  \lambda_6 \ = \ \lambda_7\ =\ 0 \; ,
\label{so5}
\end{align} 
Thus, the 2HDM potential~\eqref{pot} is parametrized by just a {\em single} mass parameter $\mu_1^2=\mu_2^2\equiv \mu^2$ and a {\em single} quartic coupling $\lambda_1=\lambda_2=\lambda_3/2\equiv \lambda$, as in the SM: 
 \begin{align}
   \label{VSO5}
V \ & = \  -\,\mu^2\, \Big(|\Phi_1|^2+|\Phi_2|^2\Big)\: +\: \lambda\,
\Big(|\Phi_1|^2+|\Phi_2|^2\Big)^2 
 \  = \ -\: \frac{\mu^2}{2}\, {\bm \Phi}^\dagger\, {\bm \Phi}\ +\ 
\frac{\lambda}{4}\, \big( {\bm \Phi}^\dagger\, {\bm \Phi}\big)^2   \; .
\end{align}
%
%
Given the 
isomorphism of the Lie algebras  ${\rm SO(5)}  \sim  {\rm Sp}(4)$, the maximal symmetry group  
of the 2HDM in      the      original      {$\bm \Phi$}-field      space
is ${\rm G}^{\bm \Phi}_{\rm 2HDM} = \left[{\rm Sp}(4)/Z_2\right] \times
{\rm SU(2)}_L$~\cite{Pilaftsis:2011ed, Dev:2014yca} in the custodial symmetry limit of vanishing  $g'$ and fermion Yukawa couplings. 

\subsection{Scalar Spectrum}\label{sec:spec-ms}
Using the parameter relations given by \eqref{so5}, we find from~\eqref{mass1} and \eqref{mass3} that in the MS-2HDM, the CP-even scalar $H$ has mass $M_H^2=2\lambda_2
v^2$, whilst  the remaining four  scalar fields, denoted  hereafter as
$h$,  $a$ and  $h^\pm$, are  massless. This is a consequence of the Goldstone theorem, since after  electroweak symmetry  breaking, ${\rm  SO}(5) \  \xrightarrow{\langle \Phi_{1,2}\rangle \neq 0}  \ {\rm
  SO}(4)$. Thus, we identify
$H$ as  the SM-like Higgs  boson with the mixing  angle $\alpha=\beta$
[cf.~(\ref{HSM})], i.e. the SM alignment limit can be naturally attributed to the SO(5) symmetry of the theory.  

In the exact SO(5)-symmetric limit, the scalar spectrum of the MS-2HDM
is experimentally unacceptable. This is because the four massless pseudo-Goldstone
particles, viz.~$h$,  $a$ and $h^\pm$,  have sizable couplings to  the 
SM $Z$ and $W^\pm$ bosons, and could induce additional decay channels,
such as~$Z\to  ha$ and $W^\pm  \to h^\pm h$, which  are experimentally
excluded~\cite{PDG}. However, as we will see below,
the   SO(5)  symmetry  may   be  violated
predominantly by  RG effects due to $g'$  and third-generation Yukawa
couplings, as well as by soft SO(5)-breaking mass parameters, thereby 
lifting the masses of these pseudo-Goldstone particles to be consistent with the experimental constraints.

\subsection{RG and Soft Breaking Effects \label{sec:RGE}}
\begin{figure}[t]
\centering
\includegraphics[width=6cm]{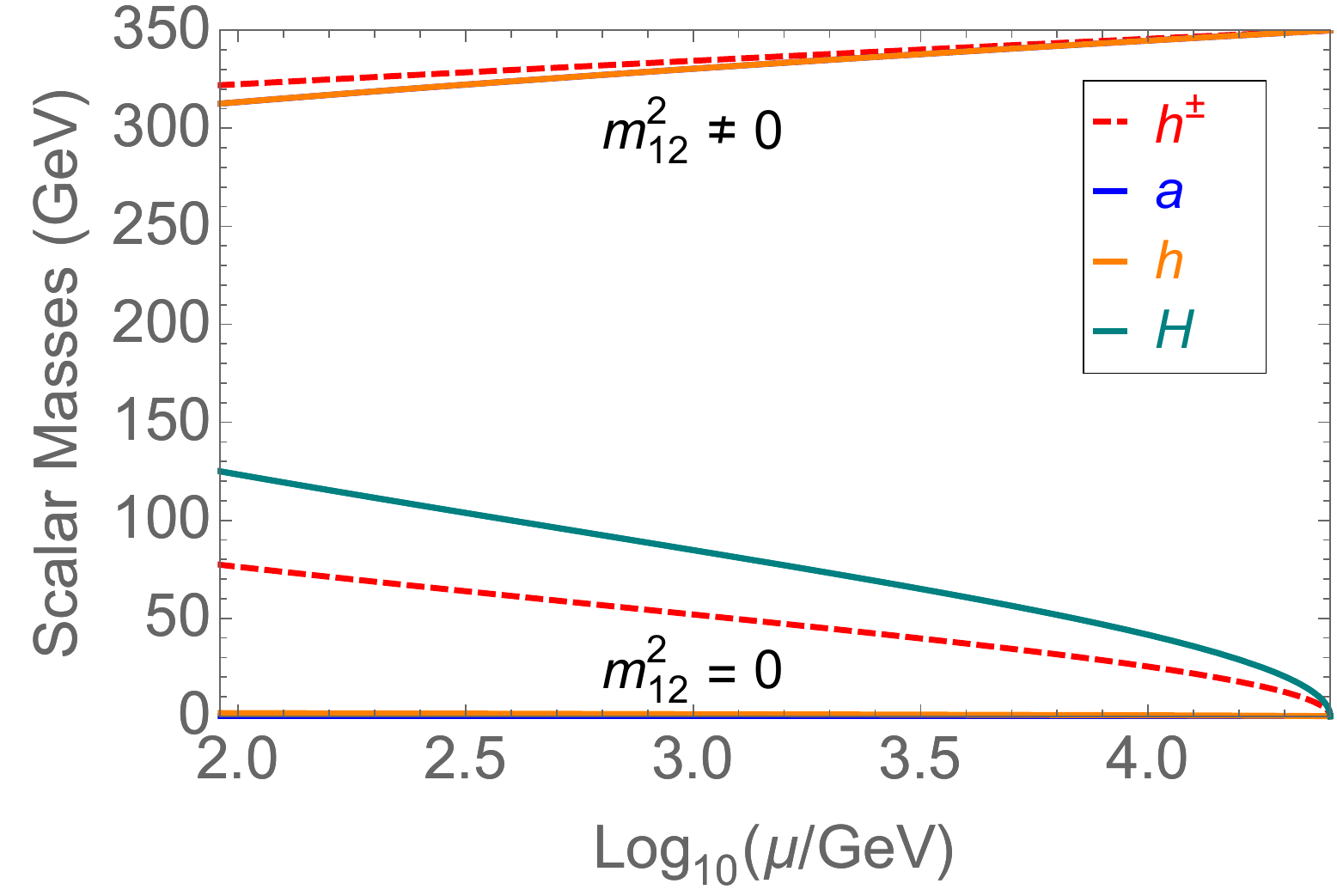}
\caption{The scalar spectrum  in the MS-2HDM  without and
  with soft-breaking effects.  
} \label{fig1}
\end{figure}

To calculate the RG and soft-breaking effects in a technically natural manner, we assume that the  SO(5)  symmetry is  realized  at  some  high scale~$\mu_X$ much above the electroweak scale. 
The
physical mass  spectrum at the  electroweak scale is then  obtained by
the RG evolution  of the 2HDM parameters given  by (\ref{pot}).  Using
the two-loop RG equations (RGEs) given in Appendix~\ref{app:RGE}, 
we  first examine the
deviation of the Higgs spectrum  from the SO(5)-symmetric limit due to
$g'$   and  Yukawa   coupling   effects, in the absence of the soft-breaking term.   This   is  illustrated   in
Figure~\ref{fig1}  for a  typical choice  of parameters  in  the Type-II
realization of the  2HDM. We find that the RG-induced $g'$ effects only lift the charged Higgs mass
$M_{h^\pm}$, while the corresponding Yukawa coupling effects also lift
slightly  the  mass of  the  non-SM CP-even  pseudo-Goldstone
boson~$h$.  However, they still leave the CP-odd scalar $a$ massless,  which can be identified  as a
${\rm  U}(1)_{\rm PQ}$  axion~\cite{Peccei:1977hh}.  

Therefore, $g'$ and Yukawa coupling effects are
{\em  not} sufficient to  yield a  viable Higgs  spectrum at  the weak
scale, starting  from a  SO(5)-invariant boundary condition~\eqref{so5}  at some  high scale
$\mu_X$.   To minimally circumvent  this problem,  we include
soft SO(5)-breaking  effects, by assuming  a non-zero soft-breaking term ${\rm
  Re}(m_{12}^2)$.   
In the  SO(5)-symmetric limit  for the
scalar  quartic couplings,  but  with ${\rm  Re}(m_{12}^2)\neq 0$,  we
obtain the following mass spectrum [cf.~(\ref{mass1}) and \eqref{mass3}]:
\begin{eqnarray}
M_H^2 \ = \ 2\lambda_2 v^2\; , \qquad M_h^2 \ = \ M_a^2 \ = \ M^2_{h^\pm} \ = \
\frac{{\rm Re}(m^2_{12})}{s_\beta c_\beta} \; ,
\label{mass-so5}
\end{eqnarray}
as  well as  an equality  between  the CP-even  and CP-odd  mixing
angles: $\alpha = \beta$, thus predicting an {\it exact} alignment for
the  SM-like  Higgs  boson  $H$, simultaneously with  
an experimentally allowed heavy Higgs spectra (cf. Figure~\ref{fig1} for $m^2_{12}\neq 0$ case). Note that in the alignment limit, the heavy Higgs sector is exactly degenerate [cf.~(\ref{mass-so5})] at the SO(5) symmetry-breaking scale, and at the low-energy scale, this degeneracy is mildly broken by the RG effects. Thus, we obtain a {\em quasi-degenerate} heavy Higgs spectrum, which is a unique prediction of the MS-2HDM, valid even in the non-decoupling limit, and can be used to distinguish this model from other 2HDM scenarios. 

\subsection{Misalignment Predictions \label{sec:MP}}
\begin{figure}[t!]
\centering
\includegraphics[width=6cm]{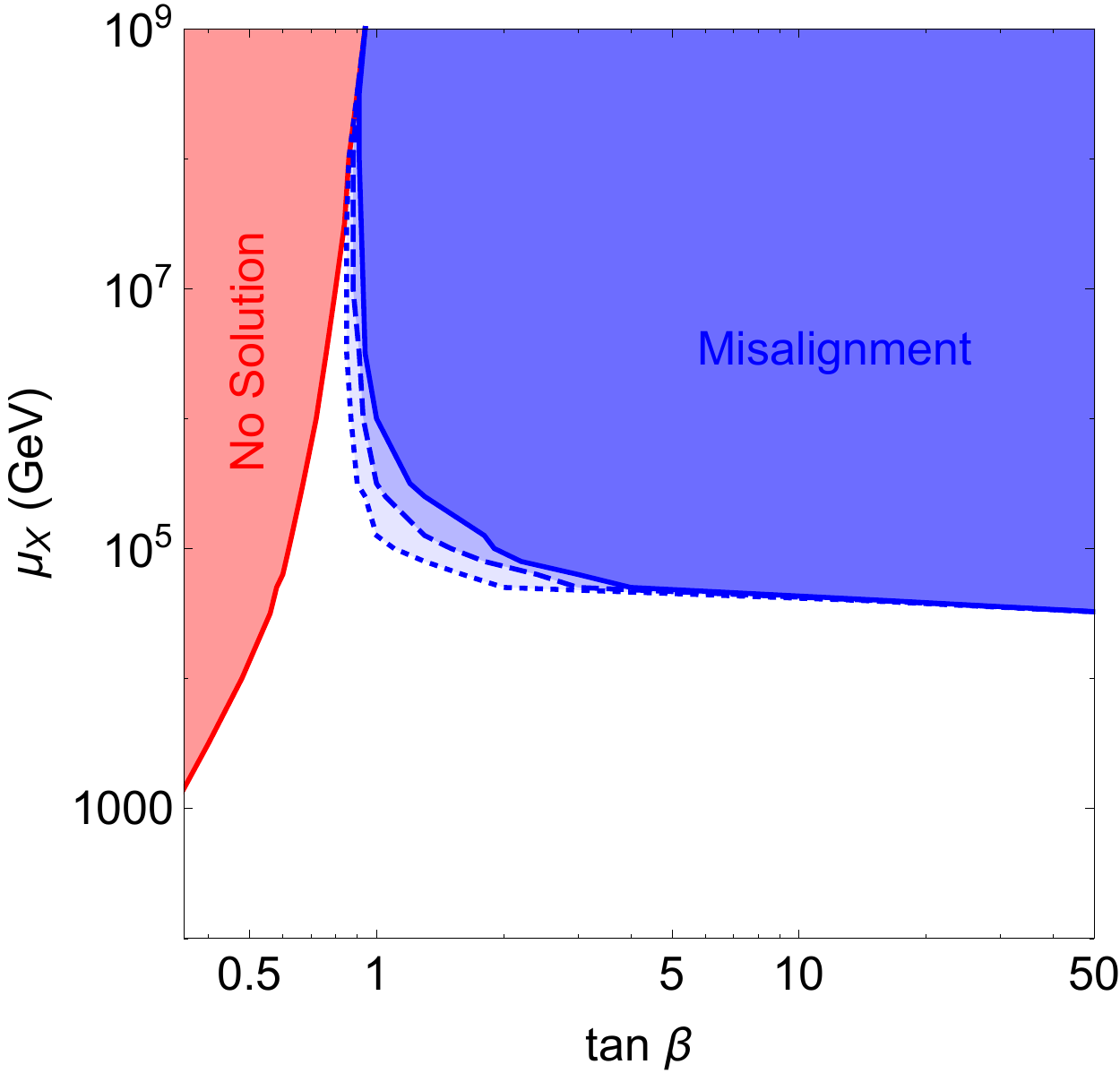}
\caption{The 
$1\sigma$ (dotted), $2\sigma$ (dashed) and $3\sigma$ (solid) exclusion contours (blue shaded region) 
from the alignment constraints in MS-2HDM. The red shaded region is theoretically excluded.
} \label{fig2} 
\end{figure}

As discussed in Section~\ref{sec:RGE}, there will be some deviation
from  the  alignment  limit  in  the low-energy  Higgs  spectrum of the MS-2HDM due to RG and soft-breaking effects.   By requiring that the  mass and couplings of the  SM-like Higgs boson $H$  
are consistent with the LHC Higgs  data~\cite{coup}, we derive predictions for the remaining
scalar spectrum and compare them with the existing (in)direct limits on the heavy
Higgs sector. 
We use the constraints in  the $(\tan\beta,~\beta-\alpha)$ plane derived from
a       recent       global       fit      for       the       Type-II
2HDM~\cite{Eberhardt:2013uba}, and require that for   a  given  set  of   SO(5)  boundary  conditions
$\big\{\mu_X,\tan\beta(\mu_X),\lambda(\mu_X)\big\}$, the
RG-evolved 2HDM  parameters at the  weak scale must satisfy  these 
alignment constraints  on the  lightest  CP-even Higgs  boson sector.   This puts stringent
constraints   on   the   MS-2HDM   parameter  space,   as   shown   in
Figure~\ref{fig2} by the blue shaded
region.    
In the red  shaded    region, there is no viable solution to the RGEs. 
We  ensure  that  the  remaining  allowed (white) region  satisfies  the
necessary theoretical constraints,  i.e.~positivity and vacuum  stability of the
Higgs     potential,    and     perturbativity     of    the     Higgs
self-couplings~\cite{review}.   From Figure~\ref{fig2},  we  find that
there  exists  an {\em upper}  limit  of  $\mu_X\lesssim  10^9$ GeV  on  the
SO(5)-breaking  scale   of  the   2HDM  potential,  beyond   which  an
ultraviolet completion  of the theory must be  invoked. 
The situation can be alleviated with the other two natural alignment scenarios listed in Table~\ref{tab1} and this will be the subject of a future publication. 

\begin{figure}[t!]
\centering
\includegraphics[width=6cm]{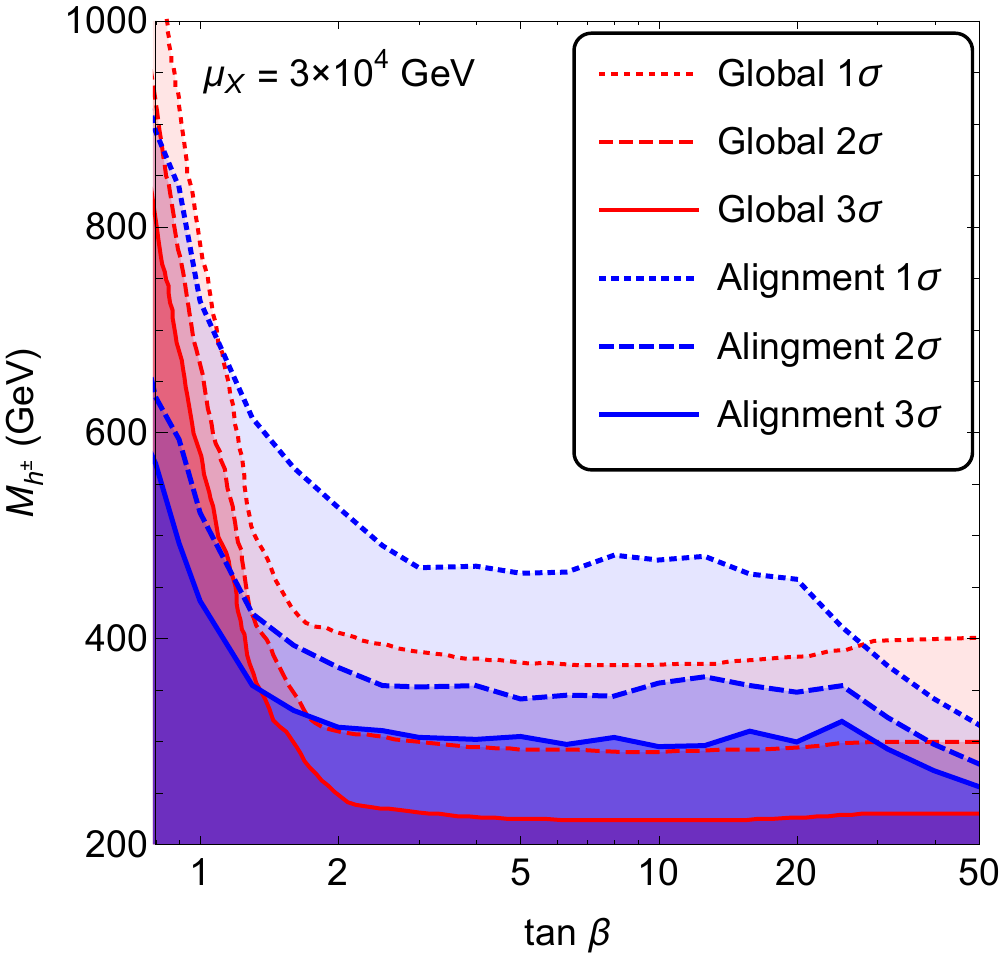}
\includegraphics[width=6cm]{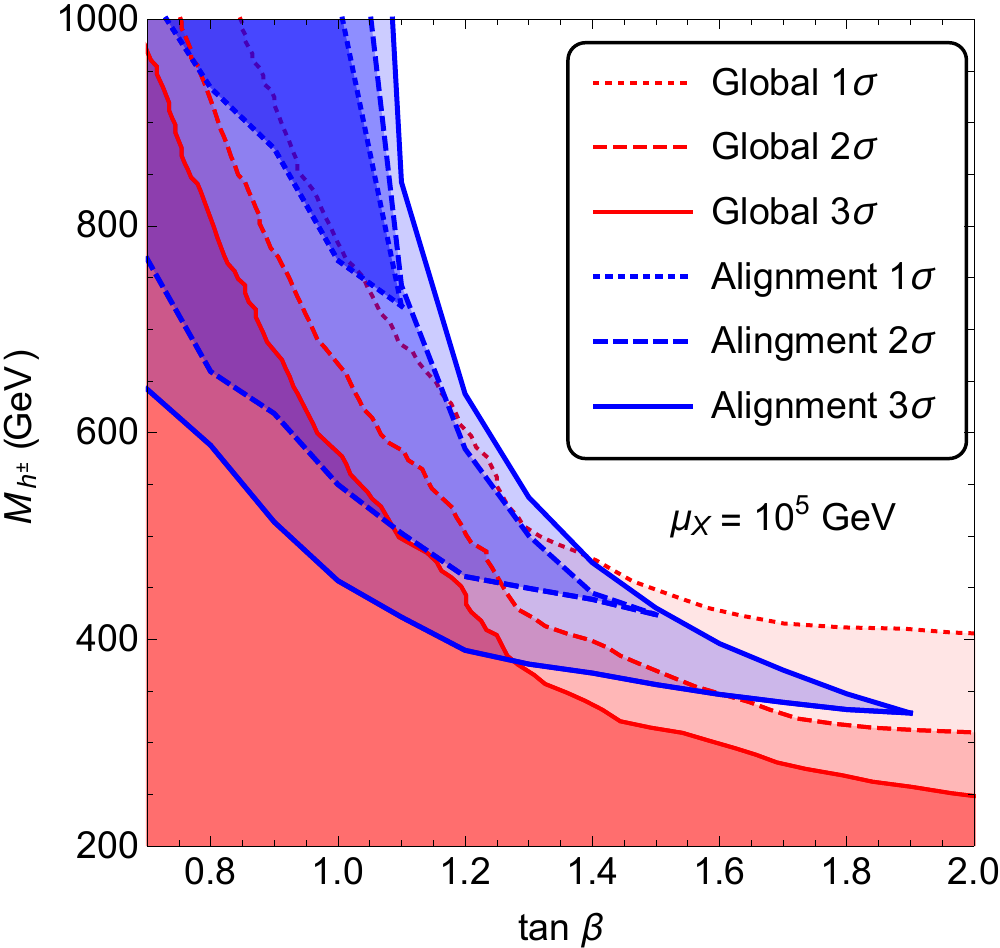}
\caption{{\em Left:} The $1\sigma$ (dotted), $2\sigma$ (dashed) and $3\sigma$ (solid) {\em lower} limits 
on the charged Higgs mass obtained from the alignment condition (blue lines) 
in the MS-2HDM with $\mu_X=3\times 10^{4}$ GeV. {\em Right:}  The $1\sigma$ (dotted), $2\sigma$ (dashed) and $3\sigma$ (solid) {\em allowed} regions from the alignment condition (blue lines) for  $\mu_X=10^{5}$ GeV. 
For comparison, the corresponding lower limits from a global fit 
are also shown (red lines). } \label{mhp}
\end{figure}

For  the  allowed   parameter  space  of  our  MS-2HDM   as  shown  in
Figure~\ref{fig2},  we obtain concrete  predictions for  the remaining
Higgs  spectrum.  In  particular,  the alignment  condition imposes  a
{\it lower} bound on the  soft breaking parameter Re$(m^2_{12})$, and hence,
on the heavy  Higgs spectrum. A  comparison of  the global fit limit on  the  charged  Higgs-boson mass  as  a function  of
$\tan\beta$~\cite{Eberhardt:2013uba} with our predicted limits from the alignment condition in the MS-2HDM for
a typical value of the boundary scale $\mu_X=3\times 10^{4}$ GeV is shown in Figure~\ref{mhp} (left panel).  It
is  clear that  the alignment  limits are  stronger than  the global fit 
limits, except  in the very small  and very large $\tan\beta$ regimes. 

From Figure~\ref{fig2}, we note that for $\mu_X\gtrsim
10^5$~GeV, phenomenologically acceptable alignment is not possible in the MS-2HDM 
for large $\tan\beta$ {\it and} large $m^2_{12}$.  Therefore, we also get an   {\it  upper}  bound  on  the  charged
Higgs-boson mass  $M_{h^\pm}$ from the misalignment condition, depending on $\tan\beta$. This is illustrated in Figure~\ref{mhp} (right panel)  
for $\mu_X=10^5$ GeV.

Similar alignment constraints are obtained  for the heavy neutral pseudo-Goldstone
bosons $h$ and  $a$, which are predicted to be quasi-degenerate with the charged Higgs boson $h^\pm$ in the MS-2HDM [cf.~(\ref{mass-so5})]. The current experimental limits on the heavy neutral Higgs sector~\cite{PDG} are weaker than the alignment constraints in this case. Thus, the MS-2HDM scenario provides a natural reason for the absence of a heavy Higgs signal below the top-quark threshold, and this has important consequences for the heavy Higgs searches in the Run-II phase of the LHC, as discussed in Sec.~\ref{sec:5}.

\section{Collider Signatures in the Alignment Limit} \label{sec:5}

In the alignment limit, the couplings of the lightest CP-even Higgs
boson are exactly similar to the SM Higgs couplings, while the heavy
CP-even Higgs boson is gaugephobic.  Therefore, two of the relevant
Higgs production mechanisms at the LHC, namely, the vector boson
fusion and Higgsstrahlung processes are suppressed for the heavy
neutral Higgs sector.  As a consequence, the only relevant production
channels to probe the neutral Higgs sector of the MS-2HDM are the
gluon-gluon fusion and $t\bar{t}h$ ($b\bar{b}h$) associated production
mechanisms at low (high) $\tan\beta$.  For the charged Higgs sector of
the MS-2HDM, the dominant production mode is the associated production
process: $gg\to \bar{t}bh^++t\bar{b}h^-$, irrespective of
$\tan\beta$. Similarly, for the decay modes of the heavy neutral Higgs
bosons in the MS-2HDM, the $t\bar{t}$ ($b\bar{b}$) channel is the
dominant one for low (high) $\tan\beta$ values, whereas for the
charged Higgs boson $h^{+(-)}$, the $t\bar{b}~(\bar{t}b)$ mode is the
dominant one for any $\tan\beta$. Thus, the heavy Higgs sector of the
MS-2HDM can be effectively probed at the LHC through the final states
involving third-generation quarks.

\subsection{Charged Higgs Signal} \label{sec5.2}
\begin{figure}[t]
\centering
\includegraphics[width=5cm]{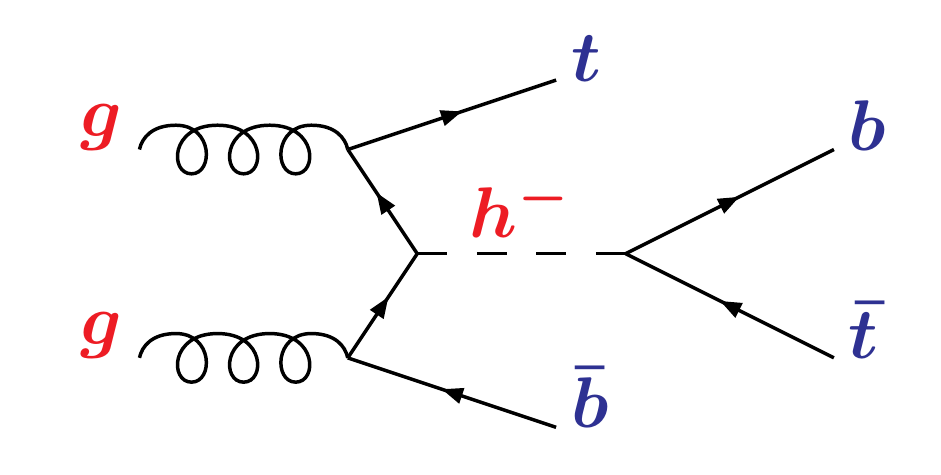} \hspace{1cm}
\includegraphics[width=6cm]{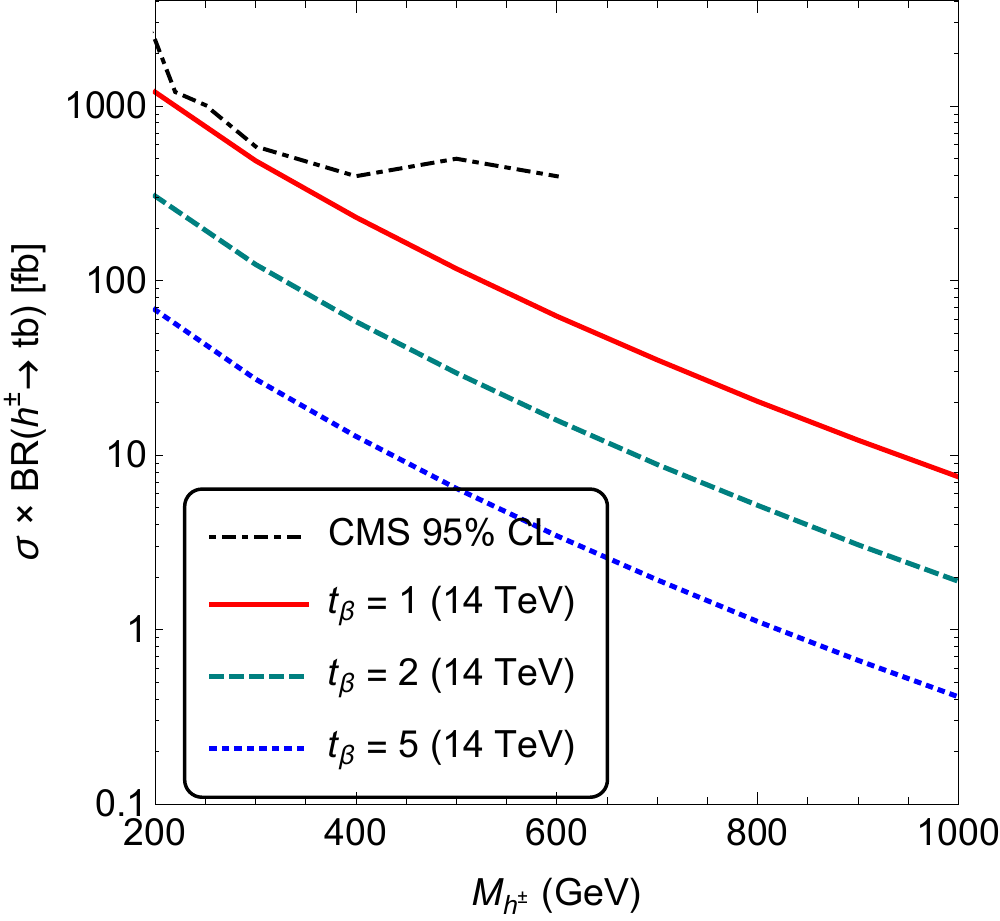}
\caption{(Left) Feynman diagram and (right) production cross sections for the charged Higgs boson in the Type-II MS-2HDM 
at $\sqrt s=14$ TeV LHC. Also shown is the 95\% CL upper limit from the $\sqrt s=8$ TeV LHC data~\cite{tb}.}    
\label{charged}
\end{figure}

The most promising charged Higgs channel in the MS-2HDM is the $t\bar{t}b\bar{b}$ final state at the LHC, as illustrated in Figure~\ref{charged} (left). 
Experimentally, this is a challenging mode due to 
large  QCD backgrounds and the non-trivial event topology, involving at least four 
$b$-jets. Nevertheless, a recent CMS study~\cite{tb} has presented a realistic analysis of this process with $\sqrt s=8$ TeV LHC data  in the leptonic decay mode of the $W$'s coming from top decay: $gg  \to h^\pm tb  \to  (\ell \nu_\ell bb)(\ell'\nu_{\ell'}b)b$  
(with $\ell,\ell'=e,\mu$). The resulting 95\% CL upper limit on the production cross section of $gg\to h^\pm tb$ times the branching ratio of $h^\pm\to tb$ is shown in Figure~\ref{charged} (right) as a function of $M_{h^\pm}$. We also show the corresponding predictions at $\sqrt s=14$ TeV LHC in the Type-II MS-2HDM for some representative values of $\tan\beta$. The cross section predictions were obtained at leading order (LO) by implementing the 2HDM Lagrangian in {\tt  MadGraph5}~\cite{mg5} and using the {\tt NNPDF2.3} parton distribution functions (PDFs)~\cite{nnpdf}. Our results in Figure~\ref{charged}  (right) suggest that the Run-II phase of the LHC could probe the low $\tan\beta$ region of the MS-2HDM parameter space using this process.  

\begin{figure}[t]
\centering
\includegraphics[width=6cm]{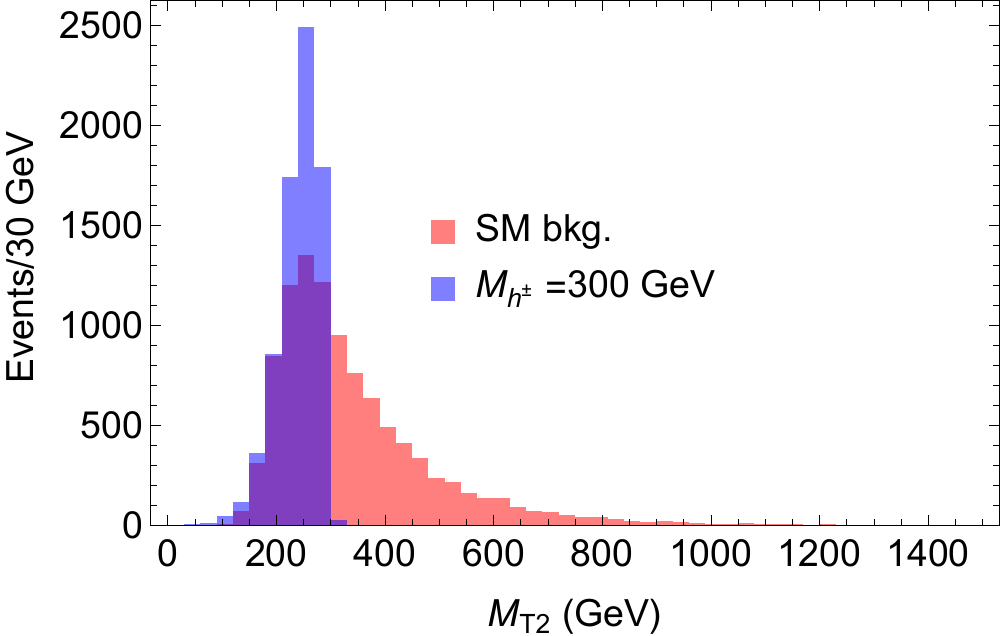} \hspace{0.5cm}
\includegraphics[width=6cm]{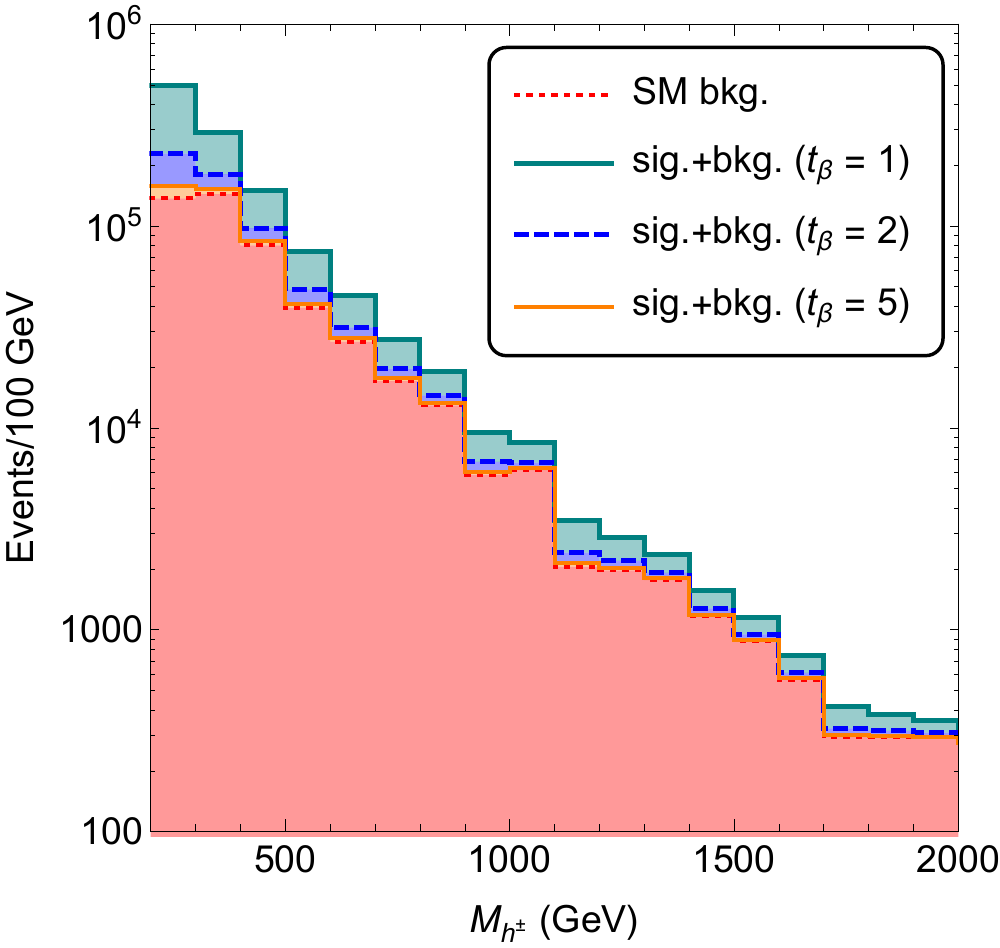}
\caption{(Left) Charged Higgs boson mass reconstruction using the $M_{T2}$ variable. (Right) The $t\bar{t}b\bar{b}$ signal in the MS-2HDM at $\sqrt
  s=14$ TeV LHC with $300~{\rm fb}^{-1}$ integrated luminosity. } 
\label{2tb}
\end{figure}
In order to make a rough estimate of the $\sqrt s=14$ TeV LHC sensitivity to the charged Higgs signal, we also perform a parton-level simulation of the $t\bar{t}b\bar{b}$ signal and background events~\cite{Dev:2014yca} (see also~\cite{Craig:2015jba}). The inclusive SM  cross section for $pp\to t\bar{t}b\bar{b}+X$  is $\sim 18$ pb at next-to-LO (NLO), with roughly 30\% uncertainty due to higher order QCD  corrections~\cite{pittau}. Most of the QCD background for the $t\bar{t}b\bar{b}$ final state can be reduced significantly by reconstructing at least one top-quark. The remaining irreducible background due to SM $t\bar{t}b\bar{b}$ production can be suppressed with respect to the signal by reconstructing the charged Higgs boson mass, once a valid signal region is defined, e.g. in terms of an observed excess of events at the LHC in future.  For the semi-leptonic decay mode of top-quarks leading to $4b+2\ell+\slashed{E}_T$ final state, one cannot directly use an invariant mass observable to infer $M_{h^\pm}$. A more useful quantity in this case is~\cite{mt2}
\begin{eqnarray}
M_{T2} \ = \ \underset{\left\{ \slashed{\mathbf p}_{T_{\rm a}}+\slashed{\mathbf p}_{T_{\rm b}}=\slashed{\mathbf p}_T\right\}}{\rm min}\Big[{\rm max}\left\{m_{T_{\rm a}},m_{T_{\rm b}}\right\}\Big] \;,
\label{mt2}
\end{eqnarray}
 where $\{{\rm a}\}, \{\rm b\}$ stand for the two sets of particles in the final state, each containing a neutrino with part of the missing transverse momentum ($\slashed{\mathbf p}_{T_{\rm {a,b}}}$). Minimization over all possible sums of these two momenta gives the observed missing transverse momentum $\slashed{\mathbf p}_T$, whose magnitude is the same as $\slashed{E}_T$ in our specific case. In (\ref{mt2}), $m_{T_{i}}$ (with $i=$a,b) is the usual transverse mass variable for the system $\{i\}$. 
For the correct combination of the final state particles, viz. $\{{\rm a}\}=(\ell \nu_\ell bb)$ and $\{{\rm b}\}=(\ell' \nu_{\ell'}bb)$ in (\ref{mt2}), the maximum value of 
$M_{T2}$ represents the charged Higgs boson mass, with the $M_{T2}$ distribution smoothly dropping to zero at this point. This is illustrated in Figure~\ref{2tb} (left) for a typical choice of $M_{h^\pm}=300$ GeV. For comparison, we also show the $M_{T2}$ distribution for the SM background, which obviously does not have a sharp endpoint. Thus, for a given hypothesized signal region defined in terms of an excess due to $M_{h^\pm}$, we may impose an additional cut on $M_{T2}\leq M_{h^\pm}$ to enhance the signal-to-background ratio.

Assuming that the charged Higgs boson mass can be reconstructed efficiently, we present an estimate of the signal and background for the charged Higgs signal in MS-2HDM at $\sqrt s=14$ TeV LHC with 300 fb$^{-1}$ for some typical values of $\tan\beta$ in Figure~\ref{2tb} (right). We find that $M_{h^\pm}$ values up to about 2 TeV can be probed via the $t\bar{t}b\bar{b}$ final state for low values of $\tan\beta$. 

\subsection{Heavy Neutral Higgs Signal} \label{sec5.3}
\begin{figure}[t!]
\centering
\includegraphics[width=5cm]{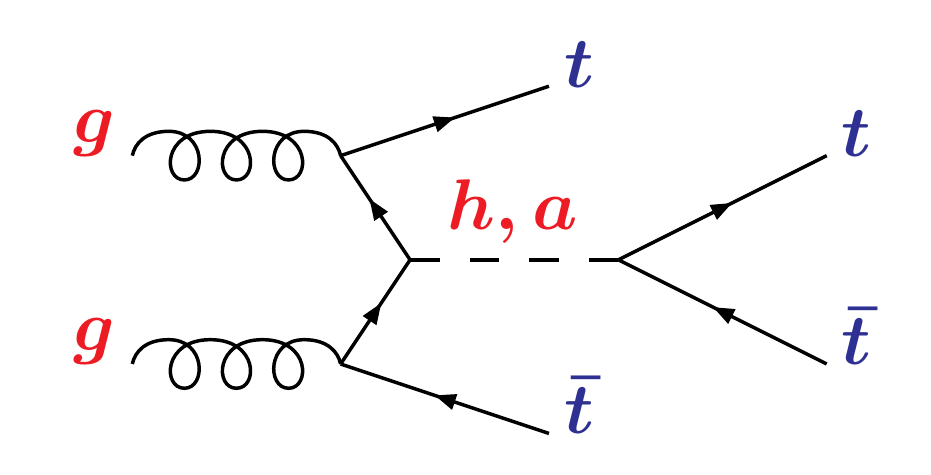} \hspace{1cm}
\includegraphics[width=7cm]{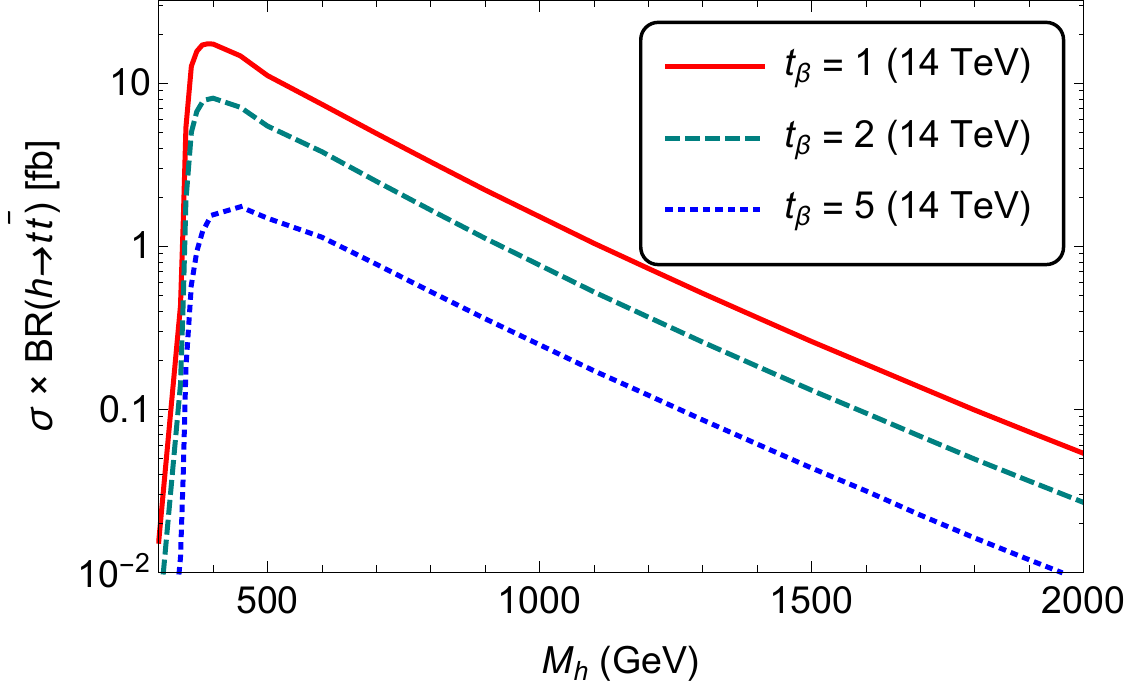}
\caption{(Left) Feynman diagram and (right) four-top production cross section in the Type-II MS-2HDM at $\sqrt s=14$ TeV LHC for various values of $\tan\beta$. } \label{4tx}
\end{figure}

So far there have been no direct searches for heavy neutral Higgs bosons involving $t\bar{t}$ and/or $b\bar{b}$ final states, mainly due to the challenges associated with uncertainties in the jet energy scales and the combinatorics arising from complicated multiparticle final states in a busy QCD environment. Nevertheless, these channels become pronounced in the MS-2HDM scenario, and hence, we have made a first attempt to study them in~\cite{Dev:2014yca}. In particular, we focus on the search channel  $gg \to  t\bar{t}h  \to t\bar{t}t\bar{t}$, as shown in Figure~\ref{4tx} (left). A more sophisticated analysis of this four-top signal, e.g. including hadron-level simulation with jet clustering, flavo and top tagging with jet substructure, detector acceptance and momentum
resolution effects at the LHC can be found in~\cite{Kanemura:2015nza}. 

To get a rough estimate of  the signal to background ratio for our 
four-top signal in the MS-2HDM,  we perform a  parton-level simulation of  the signal
and   background  events   at  LO  in   QCD   using  {\tt
  MadGraph5}~\cite{mg5}   with   {\tt NNPDF2.3}  PDF sets~\cite{nnpdf}.  For the  inclusive  SM  cross  section  for  the four-top final  state at $\sqrt s=14$ TeV LHC, we obtain 11.85 fb, whereas  our proposed signal  cross sections are
found   to  be   comparable  or   smaller  depending   on   $M_h$  and
$\tan\beta$, as shown in Figure~\ref{4tx} (right). However, since  we expect  one of  the  $t\bar{t}$ pairs coming from  an on-shell  $h$ decay to  have an invariant  mass around
$M_h$,  we can  use this information to significantly boost the signal over the irreducible SM background. Note that all the predicted cross sections shown in Figure~\ref{4tx} (right) are consistent with the current  experimental  upper bound~\cite{cms-4t}. 

Depending on the $W$ decay mode from $t\to Wb$, there are 35 final states for four top decays. Experimentally, the most favored channel is the semi-leptonic/hadronic final state with two same-sign isolated leptons: $gg \to  t\bar{t}h  \to (t\bar{t})(t\bar{t})  \to   
(\ell^\pm \nu_{\ell}b)(jjb)(\ell'^\pm \nu_{\ell'}b)(jjb)$. Although the branching fraction for this topology (4.19\%) is smaller than most of the other channels, the presence of two same-sign leptons in the final state allows us to reduce the large QCD background substantially, including that due to the SM production of $t\bar{t}b\bar{b}+$jets~\cite{thesis}. Therefore, we consider only this channel for our preliminary analysis. 

\begin{figure}[t]
\centering
\includegraphics[width=6cm]{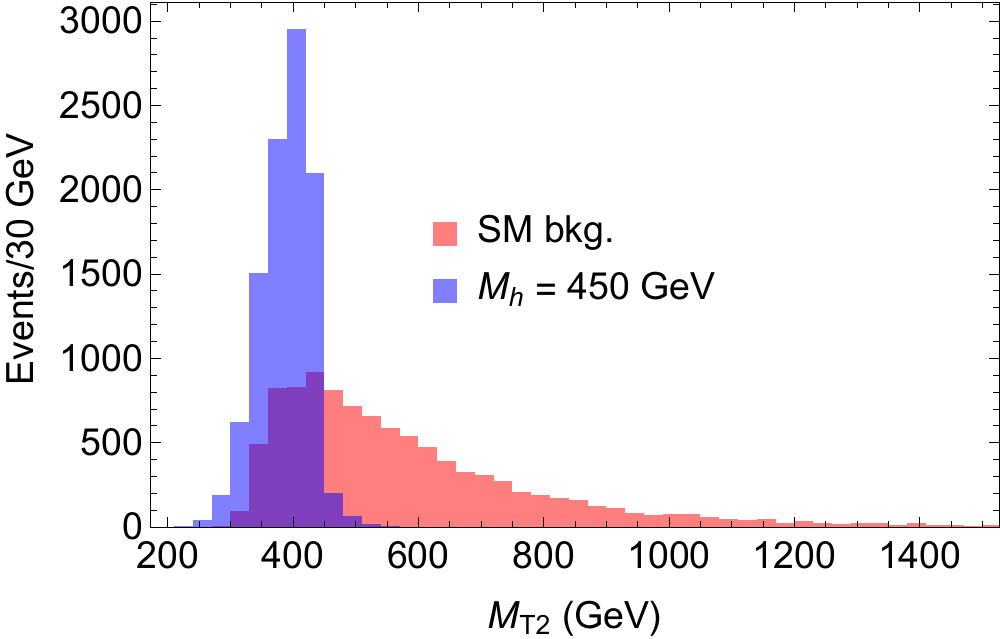} \hspace{0.5cm}
\includegraphics[width=6cm]{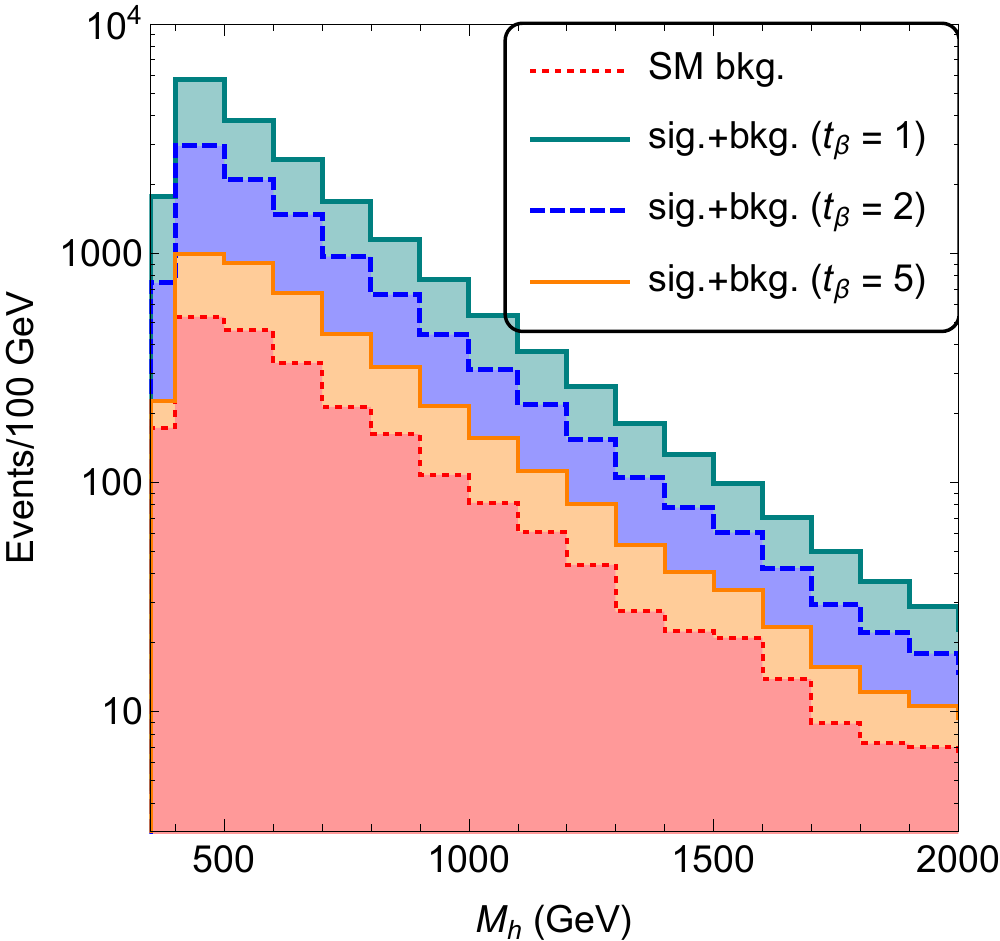}
\caption{(Left) Heavy neutral Higgs boson mass reconstruction using the $M_{T2}$ variable. (Right) The four-top signal in the MS-2HDM at $\sqrt
  s=14$ TeV LHC with $300~{\rm fb}^{-1}$ integrated luminosity.} \label{4t}
\end{figure}

As in the charged Higgs boson case, the heavy Higgs mass can be reconstructed from the signal  using the $M_{T2}$ endpoint technique, and therefore, an additional selection cut on $M_{T2}\leq M_h$ can be used to enhance the signal over the irreducible background, as illustrated in Figure~\ref{4t} (left). 
Our simulation results  for the predicted number of signal and background events at $\sqrt s=14$ TeV LHC with 300 fb$^{-1}$ luminosity are  shown in Figure~\ref{4t} (right). 
From this preliminary
analysis, we  find that the four-top channel provides the most promising collider signal
to  probe the  heavy neutral Higgs  sector in  the MS-2HDM  for low  values of
$\tan\beta \lesssim  5$.

The above analysis  is also applicable for the  CP-odd Higgs boson
$a$,  which  has  similar  production  cross  sections  and  $t\bar{t}$
branching fractions as the CP-even Higgs $h$. However, 
the $t\bar{t}h(a)$ production cross section  as well as the $h(a)\to t\bar t$ branching
ratio decreases with  increasing $\tan\beta$. This is due  to the fact
that   the   $ht\bar{t}$   coupling   in  the   alignment   limit   is
$\cos\alpha/\sin\beta\sim \cot\beta$, which is same as the $at\bar{t}$
coupling. Thus,  the high $\tan\beta$ region of
the MS-2HDM  cannot be searched  via the $t\bar{t}t\bar{t}$ channel  proposed above,
and one  needs to consider  the channels involving  down-sector Yukawa
couplings, e.g. $b\bar{b}b\bar{b}$ and $b\bar{b}\tau^+\tau^-$. It is also worth commenting here that the simpler process  $pp\to h/a \to t\bar{t}~(b\bar{b})$ at low (high) $\tan\beta$ suffers from a huge SM $t\bar{t}$ ($b\bar{b}$) QCD background, even after imposing an $M_{t\bar{t}~(b\bar{b})}$ cut. A combination of kinematic reconstruction and multivariate techniques can be 
used here to distinguish between the small signal and large background. Some recent studies on extracting the $t\bar{t}$ signal in the context of MSSM have been performed in~\cite{Djouadi:2013vqa}.


\section{Conclusions}\label{sec:6}

We provide a symmetry justification of the so-called SM alignment limit, independently of
the  heavy  Higgs  spectrum  and  the  value  of  $\tan\beta$ in the  2HDM.  
We show that there exist {\em only} three different symmetry realizations, which could lead to the SM alignment by satisfying the {\em natural alignment condition}~(\ref{alcond}) for {\em any} value of $\tan\beta$. In the context of the Maximally Symmetric 2HDM  based  on the  SO(5)  group,  we demonstrate how small
deviations from this alignment limit are naturally induced by RG 
effects  due  to  the   hypercharge  gauge  coupling  $g'$  and  third
generation Yukawa  couplings, which also break  the custodial symmetry
of  the theory.   In  addition, a  non-zero  soft SO(5)-breaking  mass
parameter is required to yield a viable Higgs spectrum consistent with
the  existing experimental constraints.   Using the LHC Higgs
data, which disfavor  large deviations
from the alignment limit, we  derive important constraints on the 2HDM
parameter space.  In particular, we  predict lower limits on the heavy
Higgs spectrum,  which prevail the present  global fit limits in a  wide range of
parameter space.   Depending on the  scale where the  maximal symmetry
could be  realized in  nature, we  also obtain an  upper limit  on the
heavy Higgs masses in certain  cases, which could be probed
during  the  Run-II phase  of  the LHC.   Finally,  we  have studied the collider signatures of the heavy Higgs sector beyond the top-quark threshold in the alignment limit. We emphasize that the final states involving third-generation quark final states can
become a valuable observational tool to directly probe the heavy Higgs
sector of the 2HDM in the alignment limit.


\section*{Acknowledgments}
This work of P.S.B.D. and A.P. is  supported   by  the   Lancaster-Manchester-Sheffield
Consortium for Fundamental Physics under STFC grant ST/L000520/1. P.S.B.D. is also supported in part by a TUM University Foundation Fellowship and the DFG cluster of excellence ``Origin and Structure of the Universe".

\appendix
\section{Two-loop RGEs}\label{app:RGE}

The  two-loop  RGEs for  the evolution of the parameters of the general  2HDM potential~\eqref{pot} are given in~\cite{Dev:2014yca}, obtained using the generic prescription given in~\cite{Machacek:1983tz}, as implemented in the public {\tt Mathematica} package {\tt  SARAH}~\cite{sarah}. Here we correct some typesetting errors in (B.9)-(B.16) of Ref.~\cite{Dev:2014yca}.\footnote{We thank Gabriel Lee and Carlos Wagner for pointing these out and for carefully checking our expressions against theirs~\cite{Lee:2015uza}.} We clarify that the numerical results presented in~\cite{Dev:2014yca} remain unchanged, as they were obtained with the correct RGEs, as directly given by the {\tt  SARAH} output.  

We start with the two-loop RGEs for the $SU(3)_c$, $SU(2)_L$ and $U(1)_Y$ gauge couplings: 
\begin{align}
{\mathcal D}g_3  \ = \ & -\frac{7
  g_3^3}{16\pi^2}\ +\ \frac{g_3^3}{256\pi^4}
\left(-26 g_3^2 + \frac{9}{2}g_2^2 + \frac{11}{6}g'^2-2y_b^2-2y_t^2
  \right)\; ,\label{g3} \\  
{\mathcal D}g_2  \ = \ & -\frac{3 g_2^3}{16\pi^2
}\ +\ \frac{g_2^3}{256\pi^4}
\left( 12g_3^2+8g_2^2+2g'^2-\frac{3}{2}y_b^2-\frac{3}{2}y_t^2-\frac{1}{2}y_\tau^2
  \right)\;,\label{g2}\\ 
{\mathcal D}g' \ = \ &  \frac{7 g'^3}{16\pi^2}\  +\
\frac{g'^3}{256\pi^4}
\left( \frac{44}{3}g_3^2+6g_2^2+\frac{104}{9}g'^2-\frac{5}{6}y_b^2
- \frac{17}{6}y_t^2  - \frac{5}{2}y_\tau^2 \right)\; , \label{g1} 
\end{align}
where ${\mathcal D}\equiv d/d\ln \mu$ ($\mu$ being the usual  't-Hooft mass employed in the regularization of  ultraviolet  divergences  in  loop  integrals), and in the two-loop terms, we have only kept the dominant third-generation contributions.

Similarly for the Yukawa RGEs, we will only consider the third-generation Yukawa
couplings, such that for Type-II 2HDM, we have
\begin{align}
{\mathcal D}y_t \  = \ & \frac{y_t}{16\pi^2 }\left(-8g_3^2-\frac{9}{4}g_2^2-\frac{17}{12}g'^2 + \frac{9}{2} y_t^2+\frac{1}{2}y_b^2 \right) \nonumber\\ 
& + \frac{y_t}{256\pi^4}\left[-108g_3^4-\frac{21}{4}g_2^4+\frac{1267}{216}g'^4+9g_2^2g_3^2-\frac{3}{4}g_2^2g'^2+\frac{19}{9}g_3^2g'^2 +6\lambda_2^2 +\lambda_3^2
 \right. \nonumber \\ 
& \left.+\lambda_3\lambda_4 +\lambda_4^2 +\frac{3}{2}(\lambda_5^2+\lambda_6^2+3\lambda_7^2) 
+\left(\frac{16}{3}g_3^2+\frac{33}{16}g_2^2 -\frac{41}{144}g'^2-2\lambda_3+2\lambda_4 \right)y_b^2 
\right. \nonumber\\
& \left.+\left(36g_3^2+\frac{225}{16}g_2^2+\frac{131}{16}g'^2-12\lambda_2\right)y_t^2 -\frac{5}{2}y_b^4-12y_t^4-\frac{5}{2}y_b^2y_t^2-\frac{3}{4}y_b^2y_\tau^2
\right]\; ,     \displaybreak \\ 
{\mathcal D} y_b \  = \ & \frac{y_b}{16\pi^2 }\left(-8g_3^2-\frac{9}{4}g_2^2-\frac{5}{12}g'^2 + \frac{9}{2}y_b^2+\frac{1}{2}y_t^2+y_\tau^2\right) \nonumber\\ 
& +\frac{y_b}{256\pi^4}\left[-108g_3^4-\frac{21}{4}g_2^4-\frac{113}{216}g'^4+9g_2^2g_3^2-\frac{9}{4}g_2^2g'^2+\frac{31}{9}g_3^2g'^2+6\lambda_1^2+\lambda_3^2\right.\nonumber\\
&\left. +\lambda_3\lambda_4+\lambda_4^2+\frac{3}{2}(\lambda_5^2+3\lambda_6^2+\lambda_7^2)+\left(\frac{16}{3}g_3^2+\frac{33}{16}g_2^2-\frac{53}{144}g'^2-2\lambda_3+2\lambda_4\right)y_t^2
\right.\nonumber\\
&\left. +\left(\frac{15}{8}g_2^2+\frac{25}{8}g'^2\right)y_\tau^2 
+\left(36g_3^2+\frac{225}{16}g_2^2+\frac{79}{16}g'^2-12\lambda_1 \right)y_b^2
\right.\nonumber\\
& \left. 
-12y_b^4-\frac{9}{4}y_\tau^4-\frac{5}{2}y_t^4-\frac{5}{2}y_b^2y_t^2-\frac{9}{4}y_b^2y_\tau^2
\right]\; , \\ 
{\mathcal D}y_\tau \  = \ & \frac{y_\tau}{16\pi^2}\left(-\frac{9}{4}g_2^2-\frac{15}{4}g'^2+3y_b^2+\frac{5}{2}y_\tau^2 \right)\nonumber \\ 
& +\frac{y_\tau}{256\pi^4}\left[-\frac{21}{4}g_2^4+\frac{161}{8}g'^4+\frac{9}{4}g_2^2g'^2  +6\lambda_1^2+\lambda_3^2+\lambda_3\lambda_4+\lambda_4^2\right. \nonumber \\ 
& \left. +\frac{3}{2}(\lambda_5^2+3\lambda_6^2+\lambda_7^2) +\left(20g_3^2+\frac{45}{8}g_2^2+\frac{25}{24}g'^2\right)y_b^2
\right.\nonumber\\
&\left. 
+\left(\frac{165}{16}g_2^2+\frac{179}{16}g'^2-12\lambda_1\right)y_\tau^2 
-\frac{27}{4}y_b^4-3y_\tau^4-\frac{27}{4}y_b^2y_\tau^2-\frac{9}{4}y_b^2y_t^2
\right]\; .
\end{align}

Similarly, the two-loop RGEs for the VEVs are given by 
\begin{align}
{\mathcal D}v_1 \ = \ &
\frac{v_1}{16\pi^2}\left[\frac{3}{4}(3g_2^2+g'^2)
- 3y_b^2-y_\tau^2\right]\nonumber\\ 
& +\frac{v_1}{256\pi^4}\left[\frac{435}{32}g_2^4-\frac{149}{32}g'^4-\frac{3}{16}g_2^2g'^2-6\lambda_1^2-\lambda_3^2-\lambda_3\lambda_4-\lambda_4^2
\right.\nonumber\\
&\left. -\frac{3}{2}(\lambda_5^2+3\lambda_6^2+\lambda_7^2) -\left(20g_3^2+\frac{45}{8}g_2^2+\frac{25}{24}g'^2\right)y_b^2-\left(\frac{15}{8}g_2^2+\frac{25}{8}g'^2\right)y_\tau^2 \right.\nonumber\\
& \left. +\frac{27}{4}y_b^4+\frac{9}{4}y_\tau^4+\frac{9}{4}y_b^2y_t^2\right]
-\frac{3 v_2}{512 \pi^4}
\bigg[(2\lambda_1+\lambda_{345})\lambda_6
+ (2\lambda_2+\lambda_{345})\lambda_7\bigg]\; , \\ 
{\mathcal D}v_2 \ = \ &
\frac{v_2}{16\pi^2}\left[\frac{3}{4}(3g_2^2+g'^2)-3y_t^2\right]\nonumber\\ 
& +\frac{v_2}{256\pi^4}\left[\frac{435}{32}g_2^4-\frac{149}{32}g'^4-\frac{3}{16}g_2^2g'^2-6\lambda_2^2-\lambda_3^2-\lambda_3\lambda_4-\lambda_4^2
\right. \nonumber\\
& \left. -\frac{3}{2}(\lambda^2_5+\lambda^2_6+3\lambda^2_7) 
- \left(20g_3^2+\frac{45}{8}g_2^2+\frac{85}{24}g'^2\right)y_t^2+\frac{9}{4}y_b^2y_t^2+\frac{27}{4}y_t^4
\right]\nonumber\\
& -\frac{3 v_1}{512 \pi^4}
\bigg[(2\lambda_1+\lambda_{345})\lambda_6+(2\lambda_2+\lambda_{345})\lambda_7
\bigg]\; .
\end{align}

The  two-loop RGEs  for all the scalar  quartic couplings appearing in (\ref{pot}) in the
Type-II 2HDM are given by
\begin{align}
{\mathcal D}\lambda_1 \ = \ & \frac{1}{16\pi^2 }\Bigg[\frac{3}{8}(3g_2^4+g'^4+2g_2^2g'^2) - 3\lambda_1(3g_2^2+g'^2)+24\lambda_1^2 + 2\lambda_3^2 + 2\lambda_3\lambda_4 + \lambda_4^2 
\nonumber\\
& \qquad \qquad 
+ \lambda_5^2 + 12\lambda_6^2 
+ 4\lambda_1 (3y_b^2+y_\tau^2)  - 6 y_b^4 - 2y_\tau^4\Bigg] \nonumber \displaybreak \\
& 
+ \frac{1}{256\pi^4}\Bigg[\frac{1}{16}\left(291 g_2^6-101 g_2^4g'^2-191g_2^2g'^4-131g'^6\right)
\nonumber\\
& -\frac{1}{8}(51g_2^4-78g_2^2g'^2-217g'^4)\lambda_1
+\frac{5}{2}(3g_2^4+g'^4)\lambda_3+\frac{5}{4}(3g_2^4+2g_2^2g'^2+g'^4)\lambda_4
\nonumber\\
& 
+(3 g_2^2+g'^2)(36\lambda_1^2+4\lambda_3^2+4\lambda_3\lambda_4
+\lambda_4^2+18\lambda_6^2)
+g'^2(\lambda_4^2-\lambda_5^2) -312\lambda_1^3 
\nonumber\\
& 
-8\lambda_3^3-6\lambda_4^3-20\lambda_1\lambda_3\lambda_4-4(5\lambda_1+3\lambda_4)
\lambda_3^2
-4(3\lambda_1+4\lambda_3)\lambda_4^2
\nonumber\\
& 
-2(7\lambda_1+10\lambda_3+11\lambda_4)\lambda_5^2 -2(159\lambda_1+33\lambda_3+35\lambda_4+37\lambda_5)\lambda_6^2 
\nonumber\\
& 
-4(9\lambda_3+7\lambda_4+5\lambda_5)\lambda_6\lambda_7
+2(3\lambda_1-9\lambda_3-7\lambda_4-5\lambda_5)\lambda_7^2
\nonumber \\ 
& 
-\left\{\frac{9}{4}g_2^4-\frac{9}{2}g_2^2g'^2-\frac{5}{4}g'^4
-\left(\frac{45}{2}g_2^2+80g_3^2+\frac{25}{6}g'^2\right)\lambda_1 
+36(4\lambda_1^2+\lambda_6^2) \right\}y_b^2 
\nonumber\\
& 
-\Big(32g_3^2-\frac{4}{3}g'^2+3\lambda_1\Big)y_b^4
 -6(2\lambda_3^2+2\lambda_3\lambda_4+\lambda_4^2
+\lambda_5^2+6\lambda_6^2)y_t^2
\nonumber\\
& 
-\left\{\frac{3}{4}g_2^4-\frac{11}{2}g_2^2g'^2
+\frac{25}{4}g'^4-\frac{5}{2}(3g_2^2+5g'^2)\lambda_1
+12(4\lambda_1^2
+\lambda_6^2)\right\}y_\tau^2
\nonumber\\
& 
-(4g'^2+\lambda_1)y_\tau^4-9\lambda_1y_b^2y_t^2+6y_t^2y_b^4
+30y_b^6+10y_\tau^6 
\Bigg]\; , 
 \\
{\mathcal D}\lambda_2 \ =\ & \frac{1}{16\pi^2 }\Bigg[\frac{3}{8}(3g_2^4+g'^4+2g_2^2g'^2) - 3\lambda_2(3g_2^2+g'^2)+24\lambda_2^2 + 2\lambda_3^2 + 2\lambda_3\lambda_4 
\nonumber\\
& \qquad \qquad  + \lambda_4^2 + \lambda_5^2 + 12\lambda_7^2 
 + 12\lambda_2 y_t^2 - 6 y_t^4\Bigg]\nonumber\\
& +\frac{1}{256\pi^4}\Bigg[\frac{1}{16}\left(291g_2^6-101g_2^4g'^2-191g_2^2g'^4-131g'^6\right)
\nonumber\\
& 
-\frac{1}{8}\left(51g_2^4-78g_2^2g'^2-217g'^4  \right)\lambda_2
+\frac{5}{2}(3g_2^4+g'^4)\lambda_3 +\frac{5}{4}(3g_2^4+2g_2^2g'^2+g'^4)\lambda_4
\nonumber\\
& 
+(3g_2^2+g'^2)(36\lambda_2^2+4\lambda_3^2+4\lambda_3\lambda_4+\lambda_4^2+18\lambda_7^2)
+g'^2(\lambda_4^2-\lambda_5^2)
-312\lambda_2^3 
\nonumber \\ 
& 
-8\lambda_3^3 -6\lambda_4^3-20\lambda_2\lambda_3\lambda_4-4(5\lambda_2+3\lambda_4)\lambda_3^2
-4(3\lambda_2+4\lambda_3)\lambda_4^2
\nonumber\\
& 
-2(7\lambda_2+10\lambda_3+11\lambda_4)\lambda_5^2 +2(3\lambda_2-9\lambda_3-7\lambda_4-5\lambda_5)\lambda_6^2
\nonumber\\
& 
-4(9\lambda_3+7\lambda_4+5\lambda_5)\lambda_6\lambda_7
-2(159\lambda_2+33\lambda_3+35\lambda_4+37\lambda_5)\lambda_7^2 
\nonumber\\
& 
-6(2\lambda_3^2+2\lambda_3\lambda_4+\lambda_4^2+\lambda_5^2+6\lambda_7^2)y_b^2
-\left\{\frac{9}{4}g_2^4-\frac{21}{2}g_2^2g'^2+\frac{19}{4}g'^4 \right.
\nonumber \\
& \left.
-\left(\frac{45}{2}g_2^2+80g_3^2+\frac{85}{6}g'^2\right)\lambda_2+36(4\lambda_2^2
+\lambda_7^2)\right\}y_t^2-9\lambda_2y_b^2y_t^2+6y_b^2y_t^4
\nonumber\\
&
-\left(32g_3^2+\frac{8}{3}g'^2+3\lambda_2\right)y_t^4+30y_t^6-2(2\lambda_3^2+2\lambda_3\lambda_4+\lambda_4^2+\lambda_5^2+
6\lambda_7^2)y_\tau^2\Bigg]\; , \\ 
{\mathcal D}\lambda_3 \ = \ & \frac{1}{16\pi^2 }\Bigg[\frac{3}{4}(3g_2^4+g'^4-2g_2^2g'^2)
-3\lambda_3(3g_2^2+g'^2)+4(\lambda_1+\lambda_2)(3\lambda_3+\lambda_4) + 4\lambda_3^2
\nonumber \\
& \quad + 2(\lambda_4^2 + \lambda_5^2) + 4(\lambda_6^2+\lambda_7^2+4\lambda_6\lambda_7)+2\lambda_3(3y_b^2+y_\tau^2+3y_t^2)-12y_b^2y_t^2
\Bigg] 
\nonumber \displaybreak \\ 
&  +\frac{1}{256\pi^4}\Bigg[\frac{1}{8}\left(291g_2^6+11g_2^4g'^2+101g_2^2g'^4-131g'^6\right)
\nonumber\\
& 
+\frac{5}{2}\left(9g_2^4-2g_2^2g'^2+3g'^4 \right)(\lambda_1+\lambda_2)
-\frac{1}{8}(111g_2^4-22g_2^2g'^2-197g'^4)\lambda_3
\nonumber\\ &
+2(3g_2^2+g'^2)[12(\lambda_1+\lambda_2)\lambda_3+\lambda_3^2+\lambda_4^2]-4(\lambda_1^2+\lambda_2^2)(15\lambda_3+4\lambda_4)
\nonumber\\
& 
 -4(\lambda_1+\lambda_2)(18\lambda_3^2+7\lambda_4^2+8\lambda_3\lambda_4+9\lambda_5^2)
-12(\lambda_3^3+\lambda_4^3+g_2^2\lambda_3\lambda_4)  
\nonumber\\
&
+\left(\frac{15}{2}g_2^4-3g_2^2g'^2+\frac{5}{2}g'^4\right)\lambda_4
+4(9g_2^2+2g'^2)(\lambda_1+\lambda_2)\lambda_4
\nonumber \\ 
&
-4\lambda_3\lambda_4(\lambda_3+4\lambda_4)
-2(9\lambda_3+22\lambda_4)\lambda_5^2
-4g'^2(\lambda_4^2-\lambda_5^2)+2g'^2(\lambda_6^2+\lambda_7^2)
\nonumber\\ 
& 
-4(31\lambda_1+11\lambda_2)\lambda_6^2-4(15\lambda_3 +17\lambda_4+17\lambda_5)(\lambda_6^2 +\lambda_7^2)+4(27g_2^2+8g'^2) \lambda_6\lambda_7 
\nonumber\\
&  -8[11(\lambda_1+\lambda_2+2\lambda_3+\lambda_4)+9\lambda_5] \lambda_6\lambda_7 
 - 4 (11\lambda_1+31\lambda_2)\lambda_7^2  
\nonumber\\
&  -\left\{ \frac{1}{4}(9g_2^4-5g'^4+18g_2^2g'^2)-\big(40g_3^2+\frac{45}{4}g_2^2+\frac{25}{12}g'^2\big)\lambda_3\right\}y_b^2
\nonumber\\
& -\frac{1}{4}\Big\{3g_2^4+25g'^4+22g_2^2g'^2-5(3g_2^2+5g'^2)\lambda_3\Big\}y_\tau^2
\nonumber\\
&  -2\left\{2 \lambda_3^2 + \lambda_4^2 + 
  4 \lambda_1 (3 \lambda_3 + \lambda_4) + \lambda_5^2 + 
  4 \lambda_6^2 + 8 \lambda_6\lambda_7\right\}(3y_b^2+y_\tau^2)
\nonumber\\
& 
 -\Big\{\frac{9}{4}g_2^4+\frac{21}{2}g_2^2g'^2+\frac{19}{4}g'^4-\Big(40g_3^2+\frac{45}{4}g_2^2+\frac{85}{12}g'^2\Big)\lambda_3 
\nonumber\\
& 
+6(12\lambda_2\lambda_3+2\lambda_3^2+4\lambda_2\lambda_4+\lambda_4^2+\lambda_5^2+8\lambda_6\lambda_7+4\lambda_7^2)\Big\}y_t^2
\nonumber\\
& 
-\big(64g_3^2+\frac{4}{3}g'^2-15\lambda_3\big)y_b^2y_t^2+36y_b^4y_t^2
-\frac{9}{2}\lambda_3[3(y_b^4+y_t^4)+y_\tau^4]+36y_b^2y_t^4 
\Bigg]\; ,  \\
{\mathcal D}\lambda_4 \ = \ & \frac{1}{16\pi^2 }\Bigg[3g_2^2g'^2 -3\lambda_4(3g_2^2+g'^2) 
+4(\lambda_1+\lambda_2+2\lambda_3)\lambda_4 + 4\lambda_4^2 + 8\lambda_5^2 
\nonumber \\
& \qquad \quad + 10(\lambda_6^2+\lambda_7^2) + 4\lambda_6\lambda_7 +2\lambda_4\big\{3(y_b^2+y_t^2)+y_\tau^2\big\}+12y_b^2y_t^2 \Bigg] \nonumber \\ 
& +\frac{1}{256\pi^4} \Bigg[-14g_2^4g'^2-\frac{73}{2}g_2^2g'^4+2g_2^2g'^2\big\{5(\lambda_1+\lambda_2)+\lambda_3\big\}
 \nonumber\\
&
-\frac{1}{8}\big(231g_2^4-102g_2^2g'^2-157g'^4\big)\lambda_4 +4\big\{2g'^2(\lambda_1+\lambda_2)
-7(\lambda_1^2+\lambda_2^2+\lambda_3^2)\big\}\lambda_4
 \nonumber\\
& 
+4(9g_2^2+g'^2)\lambda_3\lambda_4-40(\lambda_1+\lambda_2)(2\lambda_3\lambda_4+\lambda_4^2) 
+2(9g_2^2+4g'^2)\lambda_4^2-28\lambda_3\lambda_4^2
 \nonumber \\ 
& +2(27g_2^2+8g'^2)\lambda_5^2-48(\lambda_1+\lambda_2+\lambda_3)\lambda_5^2-26\lambda_4\lambda_5^2
+8g'^2\lambda_6\lambda_7
 \nonumber\\
& +2\big\{27g_2^2+7g'^2-2(18\lambda_3+17\lambda_4+20\lambda_5)\big\}(\lambda_6^2+\lambda_7^2)
-4(37\lambda_1+5\lambda_2)\lambda_6^2
\nonumber\\ &
-8\{5(\lambda_1+\lambda_2+2\lambda_3+4\lambda_4)+12\lambda_5\}\lambda_6\lambda_7 
 -4(5\lambda_1+37\lambda_2)\lambda_7^2 
\nonumber  \\ 
&+\Big\{9g_2^2g'^2+\big(40g_3^2 +\frac{45}{4}g_2^2+\frac{25}{12}g'^2 \big)\lambda_4
-12[2(\lambda_1+\lambda_3)\lambda_4+\lambda_4^2+2\lambda_5^2
\nonumber 
\\ &  
+5\lambda_6^2+\lambda_6\lambda_7]\Big\}y_b^2  
+\Big\{21g_2^2g'^2
+\big(40g_3^2+\frac{45}{4}g_2^2+\frac{85}{12}g'^2\big)\lambda_4
-12[2(\lambda_2+\lambda_3)\lambda_4
\nonumber\\
& 
+\lambda_4^2
+2\lambda_5^2+\lambda_6\lambda_7+5\lambda_7^2]\Big\}y_t^2
+\Big(64g_3^2+\frac{4}{3}g'^2-24\lambda_3-33\lambda_4  \Big)y_b^2y_t^2
\nonumber 
\displaybreak \\
&
 +\Big\{11g_2^2g'^2
+\frac{5}{4}\left(3g_2^2+5g'^2  \right)\lambda_4
-4[2(\lambda_1+\lambda_3)\lambda_4+\lambda_4^2+2\lambda_5^2
\nonumber\\
&
+5\lambda_6^2+\lambda_6\lambda_7]\Big\}y_\tau^2
-\frac{9}{2}\lambda_4\big\{3(y_b^4+y_t^4)+ y_\tau^4\big\}-24y_b^2y_t^2(y_b^2+y_t^2)
\Bigg]\; , \\ 
{\mathcal D}\lambda_5 \ = \ & \frac{1}{16\pi^2 }\Bigg[- 3\lambda_5(3g_2^2+g'^2) + 4(\lambda_1+\lambda_2+2\lambda_3+3\lambda_4)\lambda_5 
  + 10(\lambda_6^2+\lambda_7^2)
\nonumber \\ &
\qquad \qquad  + 4\lambda_6\lambda_7 
+ 2\lambda_5 \big\{3(y_b^2+y_t^2)+y_\tau^2\big\}
\Bigg] \nonumber \\ 
& +\frac{1}{256\pi^4}\Bigg[-\frac{1}{8}\Big(231g_2^4-38g_2^2g'^2-157g'^4\Big)\lambda_5 
+4(9g_2^2+4g'^2)\lambda_3\lambda_5
\nonumber\\ & 
-4\Big\{g'^2(\lambda_1+\lambda_2)+7(\lambda_1^2+\lambda_2^2+\lambda_3^2)\Big\}\lambda_5
-8(\lambda_1+\lambda_2)(10\lambda_3+11\lambda_4)\lambda_5
 \nonumber \\
&
+4\{6(3g_2^2+g'^2)-19\lambda_3-8\lambda_4\}\lambda_4\lambda_5
+6\lambda_5^3-4(37\lambda_1+5\lambda_2)\lambda_6^2
\nonumber\\
&
-4(5\lambda_1+37\lambda_2)\lambda_7^2
+2\{27g_2^2+10g'^2-2(18\lambda_3+19\lambda_4+18\lambda_5)\}(\lambda_6^2+\lambda_7^2)
\nonumber\\
&
-4\{g'^2+10(\lambda_1+\lambda_2+2\lambda_3)
+22\lambda_4+42\lambda_5\}\lambda_6\lambda_7
\nonumber \\ &
+\Big\{\big(40g_3^2+\frac{45}{4}g_2^2+\frac{25}{12}g'^2\big)\lambda_5
-12[2(\lambda_1+\lambda_3)\lambda_5+3\lambda_4\lambda_5+5\lambda_6^2+\lambda_6\lambda_7]\Big\} y_b^2 
\nonumber\\ &
+\Big\{\big(40g_3^2+\frac{45}{4}g_2^2+\frac{85}{12}g'^2\big)\lambda_5-12[2(\lambda_2+\lambda_3)\lambda_5+3\lambda_4\lambda_5+\lambda_6\lambda_7+5\lambda_7^2\Big\}y_t^2
\nonumber\\ &
+\Big\{\big(\frac{15}{4}g_2^2+\frac{25}{4}g'^2\big)\lambda_5-8(\lambda_1+\lambda_3)\lambda_5-12\lambda_4\lambda_5-20\lambda_6^2-4\lambda_6\lambda_7\Big\}y_\tau^2
\nonumber \\ &
-\frac{1}{2}\lambda_5\{3(y_b^4+y_t^4)+y_\tau^4\}-33\lambda_5y_b^2y_t^2
\Bigg]\; ,  \\ 
{\mathcal D}\lambda_6 \ = \ &  \frac{1}{16\pi^2 }\Bigg[-3\lambda_6(3g_2^2+g'^2)+2(12\lambda_1+3\lambda_3+4\lambda_4)\lambda_6+2(3\lambda_3+2\lambda_4)\lambda_7 
\nonumber \\ &
\qquad \qquad + 10\lambda_5\lambda_6+2\lambda_5\lambda_7
 +3\lambda_6(3y_b^2+y_t^2+y_\tau^2)\Bigg] \nonumber \\ 
& +\frac{1}{256\pi^4}\Bigg[   
-\frac{1}{8}(141g_2^4-58g_2^2g'^2-187g'^4)\lambda_6+6(3g_2^2+g'^2)(6\lambda_1+\lambda_3)\lambda_6
 \nonumber\\
&
-6(53\lambda_1^2-\lambda_2^2)\lambda_6
-4(33\lambda_1+9\lambda_2+8\lambda_3)\lambda_3\lambda_6
+2(18g_2^2+5g'^2)\lambda_4\lambda_6
 \nonumber\\
&
-2(70\lambda_1+14\lambda_2+34\lambda_3+17\lambda_4)\lambda_4\lambda_6
+2(27g_2^2+10g'^2)\lambda_5\lambda_6
 \nonumber\\
&
-4(37\lambda_1+5\lambda_2+18\lambda_3+19\lambda_4+9\lambda_5)\lambda_5\lambda_6
-111\lambda_6^3-42\lambda_7^3
\nonumber \\ &
+\frac{5}{4}(9g_2^4+2g_2^2g'^2+3g'^4)\lambda_7
+12(3g_2^2+g'^2)\lambda_3\lambda_7-36(\lambda_1+\lambda_2+\lambda_3)\lambda_3\lambda_7
\nonumber\\
& +2(9g_2^2+4g'^2)\lambda_4\lambda_7-2\{14(\lambda_1+\lambda_2+2\lambda_3)
+17\lambda_4\}\lambda_4\lambda_7
\nonumber  \\
& -2\{g'^2+10(\lambda_1+\lambda_2+2\lambda_3)+22\lambda_4+21\lambda_5\}\lambda_5\lambda_7
-3(42\lambda_6+11\lambda_7)\lambda_6\lambda_7
 \nonumber \\ 
&
+\Big\{60g_3^2+\frac{135}{8}g_2^2+\frac{25}{8}g'^2 
-6(24\lambda_1+3\lambda_3+4\lambda_4+5\lambda_5)
 \Big\}\lambda_6y_b^2 
 \nonumber 
 \displaybreak \\ 
&
+
\Big\{20g_3^2+\frac{45}{8}g_2^2+\frac{85}{24}g'^2-6(3\lambda_3+4\lambda_4+5\lambda_5)\Big\}\lambda_6y_t^2
 \nonumber\\
& 
+\Big\{\frac{15}{8}(3g_2^2+5g'^2)-2(24\lambda_1+3\lambda_3+4\lambda_4+5\lambda_5)\Big\}\lambda_6 y_\tau^2
\nonumber\\ &
-12(3\lambda_3+2\lambda_4+\lambda_5)\lambda_7y_t^2-\frac{1}{4}(27y_t^4+33y_b^4+11y_\tau^4)\lambda_6-21\lambda_6y_b^2y_t^2
\Bigg]\; , \\
{\mathcal D}\lambda_7 \ = \ & \frac{1}{16\pi^2 }\Bigg[-3\lambda_7(3g_2^2+g'^2)+2(12\lambda_2+3\lambda_3+4\lambda_4)\lambda_7+2(3\lambda_3+2\lambda_4)\lambda_6 
 \nonumber\\
&
\qquad \qquad + 10\lambda_5\lambda_7+2\lambda_5\lambda_6+\lambda_7(3y_b^2+9y_t^2+y_\tau^2)\Bigg]\nonumber  \\ 
& +\frac{1}{256\pi^4}\Bigg[ 
\frac{5}{4}(9g_2^4+2g_2^2g'^2+3g'^4)\lambda_6+12(3g_2^2+g'^2)\lambda_3\lambda_6
\nonumber\\
& 
-36(\lambda_1+\lambda_2+\lambda_3)\lambda_3\lambda_6
+2(9g_2^2+4g'^2)\lambda_4\lambda_6-28(\lambda_1+\lambda_2+2\lambda_3)\lambda_4\lambda_6
\nonumber \\ &
-34\lambda_4^2\lambda_6-2g'^2\lambda_5\lambda_6-4\{5(\lambda_1+\lambda_2+2\lambda_3)+11\lambda_4\}\lambda_5\lambda_6-42(\lambda_5^2+\lambda_6^2)\lambda_6
\nonumber\\
& -\frac{1}{8}(141g_2^4-58g_2^2g'^2-187g'^4)\lambda_7+6\lambda_1^2\lambda_7+36(3g_2^2
+g'^2)\lambda_2\lambda_7-318\lambda^2_2\lambda_7
\nonumber\\
&
+6(3g_2^2+g'^2)\lambda_3\lambda_7-12(3\lambda_1+11\lambda_2)\lambda_3\lambda_7
-32\lambda^2_3\lambda_7
+2(18g_2^2+5g'^2)\lambda_4\lambda_7
 \nonumber\\
&-4(7\lambda_1+35\lambda_2+17\lambda_3)\lambda_4\lambda_7
-34\lambda^2_4\lambda_7
+2(27g_2^2+10g'^2)\lambda_5\lambda_7
\nonumber\\
&
-4(5\lambda_1+37\lambda_2+18\lambda_3+19\lambda_4)\lambda_5\lambda_7-36\lambda^2_5\lambda_7-33\lambda_6^2\lambda_7
\nonumber \\ &
-126\lambda_6\lambda_7^2-111\lambda_7^3-12(3\lambda_3 +2\lambda_4+\lambda_5)\lambda_6y_b^2
\nonumber \\ &
+\Big\{20g_3^2+\frac{45}{8}g_2^2+\frac{25}{24}g'^2-6(3\lambda_3+4\lambda_4+5\lambda_5)\Big\}\lambda_7y_b^2
 \nonumber\\
&
+ \Big\{60g_3^2+ \frac{135}{8}g_2^2+\frac{85}{8}g'^2-6(24\lambda_2+3\lambda_3+4\lambda_4+5\lambda_5)
 \Big\}\lambda_7 y_t^2
 \nonumber  \\
&
-4(3\lambda_3+2\lambda_4+\lambda_5)\lambda_6y_\tau^2  
+\Big\{\frac{5}{8}(3g_2^2+5g'^2)-2(3\lambda_3+4\lambda_4+5\lambda_5)\Big\}\lambda_7 y_\tau^2
\nonumber\\ &
-\frac{1}{4}(33y_t^4+27y_b^4+9 y_\tau^4)\lambda_7-21\lambda_7y_b^2y_t^2 
\Bigg]\; . 
\end{align}
The two-loop RGE for the soft mass parameter is given by 
\begin{align}
{\mathcal D}(m^2_{12}) \ = \ & \frac{1}{16\pi^2}\Bigg[ 
-\frac{3}{2}(3g_2^2+g'^2)m^2_{12}+2(\lambda_3+2\lambda_4+3\lambda_5)m^2_{12}
\nonumber \\ 
& \qquad \qquad +2(3y_b^2+3y_t^2+y_\tau^2)m^2_{12}+12(\lambda_6\mu^2_1+\lambda_7\mu^2_2) \Bigg]
\nonumber \\ 
& +\frac{1}{256\pi^4}\Bigg[
-\frac{1}{16}(243g_2^4-30g_2^2g'^2-153g'^4)m^2_{12}+3\big\{2\lambda_1^2+\lambda_2^2)+\lambda_5^2
\nonumber \\ &
+4(\lambda_6^2+\lambda_7^2)\big\}m^2_{12}
+4(3g_2^2+g'^2)(\lambda_3+2\lambda_4+3\lambda_5) m^2_{12}
\nonumber 
\\ &
-12(\lambda_1+\lambda_2)\lambda_{345} 
m^2_{12}-6(\lambda_3\lambda_4+2\lambda_3\lambda_5+2\lambda_4\lambda_5+6\lambda_6\lambda_7)m^2_{12}
\nonumber  \displaybreak  \\ &
+\Big\{20g_3^2+\frac{45}{8}g_2^2+\frac{25}{24}g'^2
-6(\lambda_3+2\lambda_4+3\lambda_5)\Big\}y_b^2m^2_{12}
\nonumber \\ &
+\Big\{20g_3^2+\frac{45}{8}g_2^2+\frac{85}{24}g'^2
-6(\lambda_3+2\lambda_4+3\lambda_5)\Big\}y_t^2m^2_{12}
\nonumber   \\ &
+\Big\{\frac{5}{8}(3g_2^2+5g'^2)
-2(\lambda_3+2\lambda_4+3\lambda_5)\Big\}y_\tau^2 m^2_{12}
\nonumber \\ &
+24(3g_2^2+g'^2)(\lambda_6\mu_1^2+\lambda_7\mu_2^2)
-72(\lambda_1\lambda_6\mu_1^2+\lambda_2\lambda_7\mu_2^2)
\nonumber \\ &
-12\lambda_{345}\{(2\lambda_6+\lambda_7)\mu_1^2+(\lambda_6+2\lambda_7)\mu_2^2\}
\nonumber \\ &
-24\{(3y_b^2+y_\tau^2)\lambda_6\mu^2_1+3y_t^2\lambda_7\mu_2^2\}
-\frac{9}{4}(3y_b^4+3y_t^4+y_\tau^4)m^2_{12}
\Bigg]
\end{align}

\end{document}